\documentclass[%
 aip,
 amsmath,amssymb,
reprint,floatfix%
]{revtex4-1}

\usepackage{graphicx}
\graphicspath{ {./Images/} }
\usepackage{gensymb}
\usepackage{xcolor}
\usepackage{dcolumn}
\usepackage{bm}
\usepackage{physics}
\usepackage{graphicx}

\usepackage[utf8]{inputenc}
\usepackage[T1]{fontenc}
\usepackage{mathptmx}

\usepackage{tikz}
\usepackage{caption}
\usepackage{subcaption}
\usepackage{tabularx}
\usepackage{hyperref}

\usepackage{natbib}

\definecolor{green}{RGB}{0,85,80}
\definecolor{orange}{RGB}{255,110,29}
\definecolor{grey}{RGB}{83, 83, 83}

\begin{document}

\preprint{}

\title[Main]{Designing High-Power, Octave Spanning Entangled Photon Sources for Quantum Spectroscopy}

\author{S. Szoke}
\affiliation{ 
Division of Engineering and Applied Sciences, California Institute of Technology, Pasadena, CA 91125, USA
}%

\author{M. He}
\author{B.P. Hickam}
\author{S.K. Cushing}
 \email[]{scushing@caltech.edu}
\affiliation{%
Division of Chemistry and Chemical Engineering, California Institute of Technology, Pasadena, CA 91125, USA
}%

\date{\today}

\begin{abstract}
Entangled photon spectroscopy is a nascent field that has important implications for measurement and imaging across chemical, biology, and materials fields. Entangled photon spectroscopy potentially offers improved spatial and temporal-frequency resolutions, increased cross sections for multiphoton and nonlinear measurements, and new abilities in inducing or measuring quantum correlations. A critical step in enabling entangled photon spectroscopies is the creation of high-flux entangled sources that can use conventional detectors, as well as provide redundancy for the losses in realistic samples. Here, we report a periodically poled, chirped, lithium tantalate platform that generates entangled photon pairs with a $\sim10^{-7}$ efficiency. For a near watt level diode laser, this results in a near $\mu$W-level flux. The single photon per mode limit that is necessary to maintain non-classical photon behavior is still satisfied by distributing this power over up to an octave-spanning bandwidth. The spectral-temporal photon correlations are observed via a Michelson-type interferometer that measures the broadband Hong-Ou-Mandel two-photon interference. A coherence time of 245 fs for a 10 nm bandwidth in the collinear case and a 62 fs for a 125 nm bandwidth in the non-collinear case is measured using a CW pump laser, and, essentially, collecting the full photon cone. We outline in detail the numerical methods used for designing and tailoring the entangled photons source, such as changing center wavelength or bandwidth, with the ultimate aim of increasing the availability of high-flux UV-Vis entangled photon sources in the optical spectroscopy community.
\end{abstract}

\maketitle

\section{\label{introduction}Introduction}

An entangled photon pair is generated by splitting a single photon into two photons via spontaneous parametric down conversion (SPDC) \cite{Burnham1970}. The quantum correlations between the two photons have several effects. First, when coherently recombined in time and space, the two entangled photons behave as the initial pump photon. In other words, the two entangled photons can be thought of as acting like a single photon in light-matter interactions \cite{Szoke2020}. As a result, two entangled photons leave the same side of a beamsplitter \cite{Hong1987} and diffract from a grating at the wavelength of the pump photon \cite{Mitev2020,Shimizu2003,Abouraddy2001}. Entangled photon pairs also linearize multiphoton and nonlinear spectroscopy, as has been well-studied in two photon absorption and fluorescence \cite{Javanainen1990,Dayan2005,Goodson2020,Raymer2013,Tabakaev2021,Li2020,Kang2020}. Entangled photons also have non-Fourier reciprocal spectral-temporal resolutions and are able to yield spatial resolutions which scale inversely with the number of correlated entangled photons \textit{N} \cite{Boto2000, Steuernagel2004, Oka2011, Oka2018,Schlawin2018}.

These properties, plus the more well known sub-shot noise and classical light rejection behaviors, make entangled photons tempting for use in ultrafast laser spectroscopy \cite{Dorfman2016,Maclean2019}. The fragility of the entangled states, and the difficulty and expense associated with single photon counting, necessitates a need for high flux (nW-$\mu$W) entangled photon sources that would easily interface with existing ultrafast optics setups \cite{Magued2005}. However, commonly used nonlinear crystals (BBO, KDP), usually have entangled photon creation efficiencies in the $10^{-10}$ to $10^{-12}$ range \cite{Kwiat1995,Friberg1985,Kwiat1999}. Reliance on 'birefringent phase matching' limits the highest usable nonlinear coefficients of the crystal as well as the bandwidth of entangled photons that can be created \cite{Karan2020, Nikogosyan1991, Midwinter1965}. The possible phase-matching conditions result in bandwidths in the tens of nanometers range, which corresponds to 100s of femtoseconds to picoseconds in correlation times in experiments. Given the current assumption that the entangled photon correlation time must be less than a molecule or material's dephasing time to allow a nonlinear or multiphoton enhancement, the narrow bandwidth of birefringent phase matching presents critical limits on entangled spectroscopy's feasibility \cite{Dorfman2016, Roslyak2009}.

In contrast, the ferroelectric properties of materials such as KTP and lithium niobate  allow the crystals to be periodically poled and utilize a 'quasi phase-matching' (QPM) technique \cite{Wang1999,Yu2002,Lin2019}. Quasi-phase-matching allows the use of Type-0 collinear phase matching \cite{Chen2009}, SPDC efficiencies of $10^{-6}$ to $10^{-10}$ \cite{Bock2016,Shi2004}, a broadly tunable center wavelength (set by the grating period instead of crystal angle) \cite{Armstrong1962}, and easy implementation of waveguides that can saturate single photon detectors with $\sim$mW of CW diode laser input power \cite{Tanzilli2001}. Broad bandwidth phase matching is achieved by introducing a chirp into the poling period of the structure. The nonlinear poling allows for more wavelengths to be phase matched, while also retaining the benefits associated with longer crystals, namely a larger downconverted photon flux. The increased bandwidth directly affects the effective temporal resolution of experiments \cite{Nasr2008,Peer2005,Ou1999}.The broad bandwidth also allows increased power densities, as the the single photon per wavelength mode occupation limit necessary for non-classical effects is not exceeded (calculated as the flux in photons/s divided by the bandwidth in Hz) \cite{Hadfield2009}. While KTP and lithium niobate have been the primary workhorse for entangled photon generation in quantum information system settings, they have limitations with regards to their damage thresholds and UV cut-offs. In contrast, lithium tantalate is more ideal for physical chemistry and materials research because it has higher power handling capabilities (240 MW/cm\textsuperscript{2}) \cite{Zverev1972} and a lower UV absorption edge ($\approx$280 nm) \cite{Antonov1975,Meyn1997}. 

To accelerate the adoption of entangled photon techniques, we outline how to create an entangled photon source that can achieve near $\mu$W powers of entangled photons down to UV wavelengths with temporal resolutions of tens of femtoseconds. These entangled photon fluxes are high enough to be visible by eye when starting from a 1W CW pump laser, greatly facilitating the alignment of entangled experiments. Here, the generated photon pairs are centered around a degenerate wavelength of 812 nm, resulting from a 406 nm pump source, but this wavelength can be tuned as outlined by the formulas in the paper. The 8\% MgO doped congruent lithium tantalate (CLT) gratings use a Type-0, collinear quasi-phase matching configuration. A Michelson type interferometric scheme \cite{LopezMago2012} measures the fourth-order interference of the entangled photons and gives coherence times of 245 fs for a bandwidth of 10 nm and 62 fs for a bandwidth of 125 nm. Of course, simply producing a broad bandwidth of entangled photons is not the same as carrying out experiments using them. We discuss our attempts to utilize the full SPDC cone and bandwidth in subsequent experiments, emphasizing where optics development is still needed to achieve maximum temporal resolution and flux. Using the full cone is in contrast to most experiments where irises are used to select two points from the SPDC cone, thereby severely reducing the total photon flux at the gain of state purity. The paper aims to give quick access to the sources needed for the emerging field of nonlinear entangled photon optics as well as outlining the next steps needed in the field.

\section{\label{theory}Theory}

\subsection{\label{spdc}Designing Broadband SPDC in chirped nonlinear materials}

Prior to quantifying the experimental behavior of the SPDC sources, we lay out the theoretical and numerical details used in their design.

\begin{figure}[ht]
\includegraphics[width=\columnwidth]{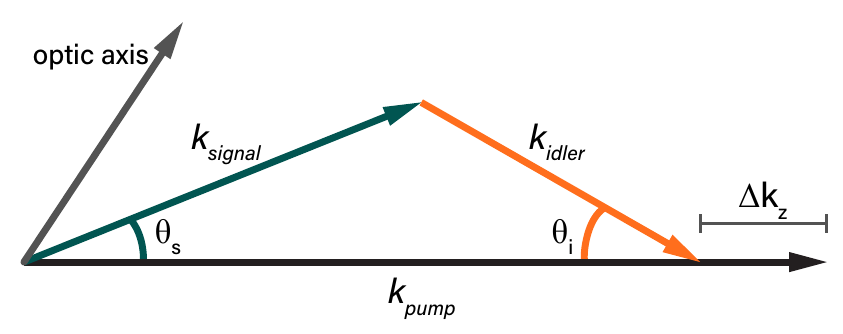}%
\caption{Phase-matching diagram for a two-photon downconversion process, showing the longitudinal phase-mismatch component $\Delta k_z$. The entangled photons (signal and idler) are each shown to have their distinct wave vectors \textit{k} and emission angles $\theta$. The phase-mismatch component must be compensated to create entangled photons by spontaneous parametric down conversion (SPDC).\label{spdc-vector}}%
\end{figure}

In SPDC, the creation of two entangled photons arises from the quantum mechanical process of a single pump photon mixing with the underlying vacuum state. The two down-converted daughter photons display strong correlations in time, energy, and momentum due to the parametric mixing process \cite{Burnham1970, Klyshko1970}. The time and energy correlations originate from the individual photons being generated simultaneously from the pump photon:
\begin{equation}
    E_p\left( \omega_p\right )=E_s\left( \omega_s\right )+E_i\left( \omega_i\right )
\end{equation}
\begin{equation}
    \therefore\;\omega_i=\omega_p-\omega_s
\end{equation}
where $\omega_p$, $\omega_s$, and $\omega_i$ are the frequencies of the pump, signal, and idler photon respectively. The momentum correlation is dictated by the phase matching condition, which is a consequence of the wave-like nature of the interacting fields. Each field has a corresponding wave vector defined via its respective phase velocity $\textit{v}_p=c/n\left( \omega,T\right )$ in the medium:
\begin{equation}
    k\left ( \omega,T \right )=\frac{\omega n\left ( \omega,T \right )}{c}
\end{equation}
with $n\left ( \omega,T \right )$ being the frequency and temperature dependent refractive index of the material. The following vector sum, the longitudinal component of which is depicted in Fig.\ref{spdc-vector}, should be satisfied in order to ensure that a proper phase relationship between the interacting waves is maintained throughout the nonlinear crystal.
\begin{equation}
    \begin{aligned}
    \Delta \textbf{\textit{k}}\left ( \omega_p,\omega_s,\omega_i \right )=\textbf{\textit{k}}_p\left ( \omega_p,n_p\left(\omega_p,T\right) \right )
    -\textbf{\textit{k}}_s\left ( \omega_s,n_s\left(\omega_s,T\right) \right )\\
    -\textbf{\textit{k}}_i\left ( \omega_i,n_i\left(\omega_i,T\right)\right )
    \end{aligned}
\end{equation}
It is only when this relationship holds true $\left(\Delta k=0\right)$ that constructive interference between the propagating waves can occur and energy is transferred dominantly in one direction, rather than oscillating back and forth between the two optical modes \cite{Boyd2008}.

A natural extension of this phase-matching principle, termed quasi-phase-matching, is apparent if one considers the duality between position and momentum space. A periodic lattice in position space directly corresponds to a well-defined wave vector $\textbf{\textit{k}}$ in momentum space via a Fourier transform. Therefore, if the nonlinear dielectric tensor domains of ferroelectric nonlinear crystal are modulated with a specific periodicity $\Lambda$, a quasi-momentum term $\Gamma\left(\Lambda \right)$ arises, which can be used to satisfy a non-phase-matched nonlinear interaction.
\begin{equation}\label{deltaKQPM}
    \Delta \tilde{\textbf{\textit{k}}}\left ( \omega_p,\omega_s,\omega_i \right )=\Delta \textbf{\textit{k}}\left ( \omega_p,\omega_s,\omega_i \right )-\Gamma\left ( \Lambda \right )
\end{equation}
The quasi-phase-matching term $\Gamma\left(\Lambda \right)$ also has an additional advantage, in that it can be chosen to be higher-\textit{order}. Here, the order of the configuration is defined by \textit{m}.
\begin{equation}
\Gamma\left ( \Lambda \right )=\frac{2\pi m}{\Lambda},\;m \in 2\mathbb{Z}+1 \left ( \neg0 \right )
\end{equation}
This serves a particular experimental benefit as it allows for nonlinear interactions to be phase-matched (albeit with lower efficiency), even if the required $1^{st}$ order periodicity is too small to be fabricated \cite{fejer1992}.

We now describe the design of a grating in congruent lithium tantalate with the design goal of a 406 nm pump wavelength that creates a >400 nm bandwidth around the degenerate 812 nm SPDC point. The source must also be capable of creating near $\mu$W powers given a 1 W input power. The first step is to obtain the Sellmeier equations that describe the frequency and temperature dependent refractive indices of the material \cite{Moutzouris2011}. The numeric values and a representative plot are found in the supplementary information. Using the Sellmeier equations, the phase mismatch (Eq.\ref{deltaKQPM}) is then evaluated for a wide range of pump-signal/idler wavelength configurations while keeping one entangled photon frequency a constant. External conditions such as the desired temperature range and minimum achievable poling periodicity are also enforced such that the parameter space remains within experimentally realistic boundaries. The root of Eq.\ref{deltaKQPM} can then be found by a root-finding algorithm. For the congruent lithium tantalate and a 406-812 SPDC process, and imposing a realistic < 200\degree C temperature condition, solving for the root of Eq.\ref{deltaKQPM} yields a poling period of 9.5 $\mu$m at 133\degree C (see figure in SI for the graphical solution).

\begin{figure}[ht]
\includegraphics[width=\columnwidth]{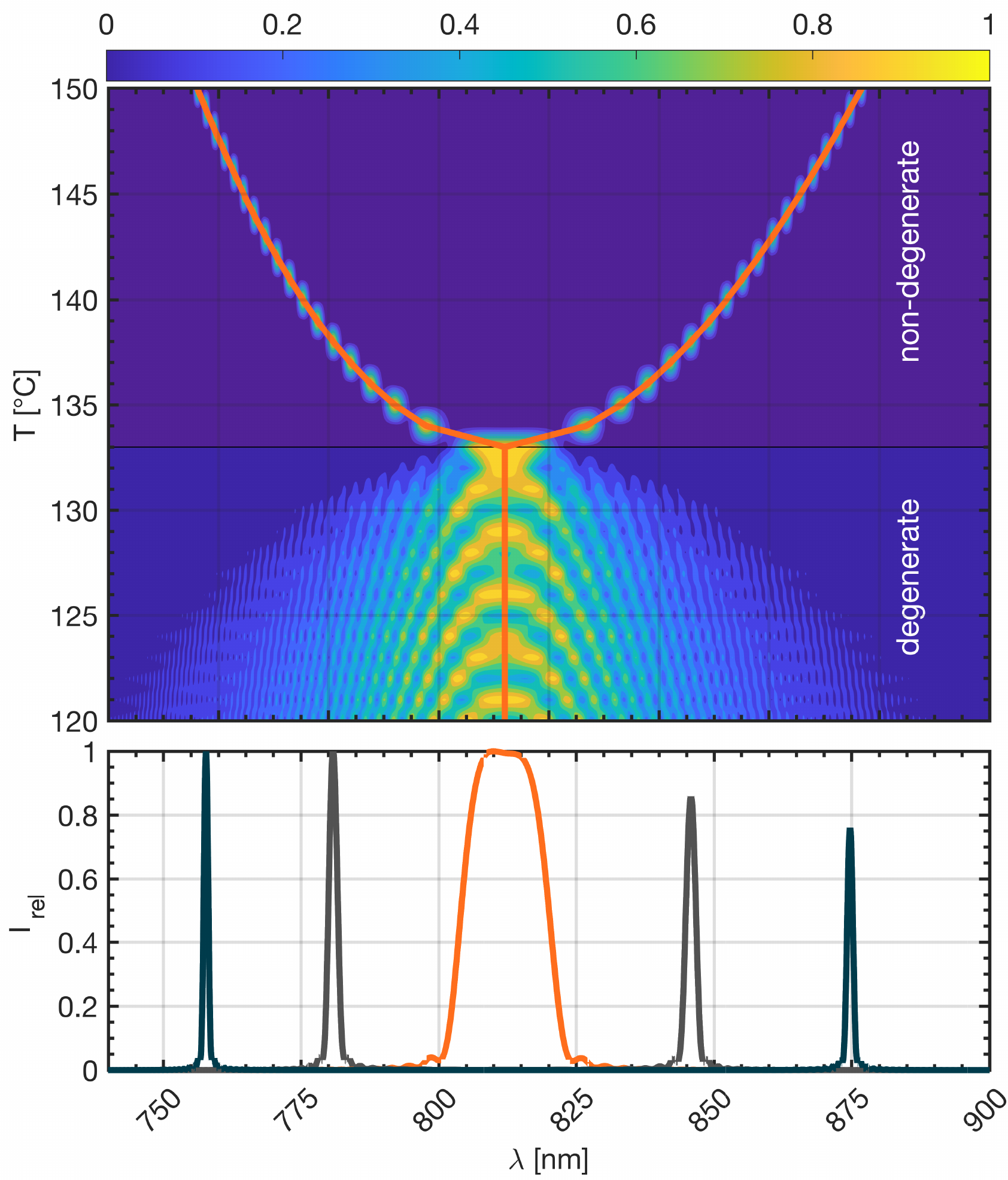}%
\caption{SPDC emission spectrum for ppCLT at various phase-matching temperatures showing transition from degenerate to non-degenerate downconversion. Bottom plot showing spectra at three temperatures corresponding to cross sections of the 2D plot. The phase matching temperature results in the broadest bandwidth.\label{ppCLT-spectrum-Temp}}%
\end{figure}

Quasi-phase-matching is a strongly temperature dependent technique. By solving Eq.\ref{deltaKQPM} for a range of temperatures near the calculated degeneracy point (133\degree C), the splitting and tuning of the SPDC spectrum can be calculated. Tuning the SPDC spectrum via the crystal temperature is useful for controlling what intermediate states are involved in, the magnitude of, and the temporal resolution of an entangled two photon process. The temperature dependent behavior is shown in Fig.\ref{ppCLT-spectrum-Temp}. Here, the transition between the degenerate and non-degenerate domains is clearly observable, with lower temperatures corresponding to a weaker degenerate emission profile. The lower plot of Fig.\ref{ppCLT-spectrum-Temp} also clearly shows a very strong narrowing of the signal and idler bandwidths as one temperature tunes further away from degeneracy.

\begin{figure}[ht]
\includegraphics[width=\columnwidth]{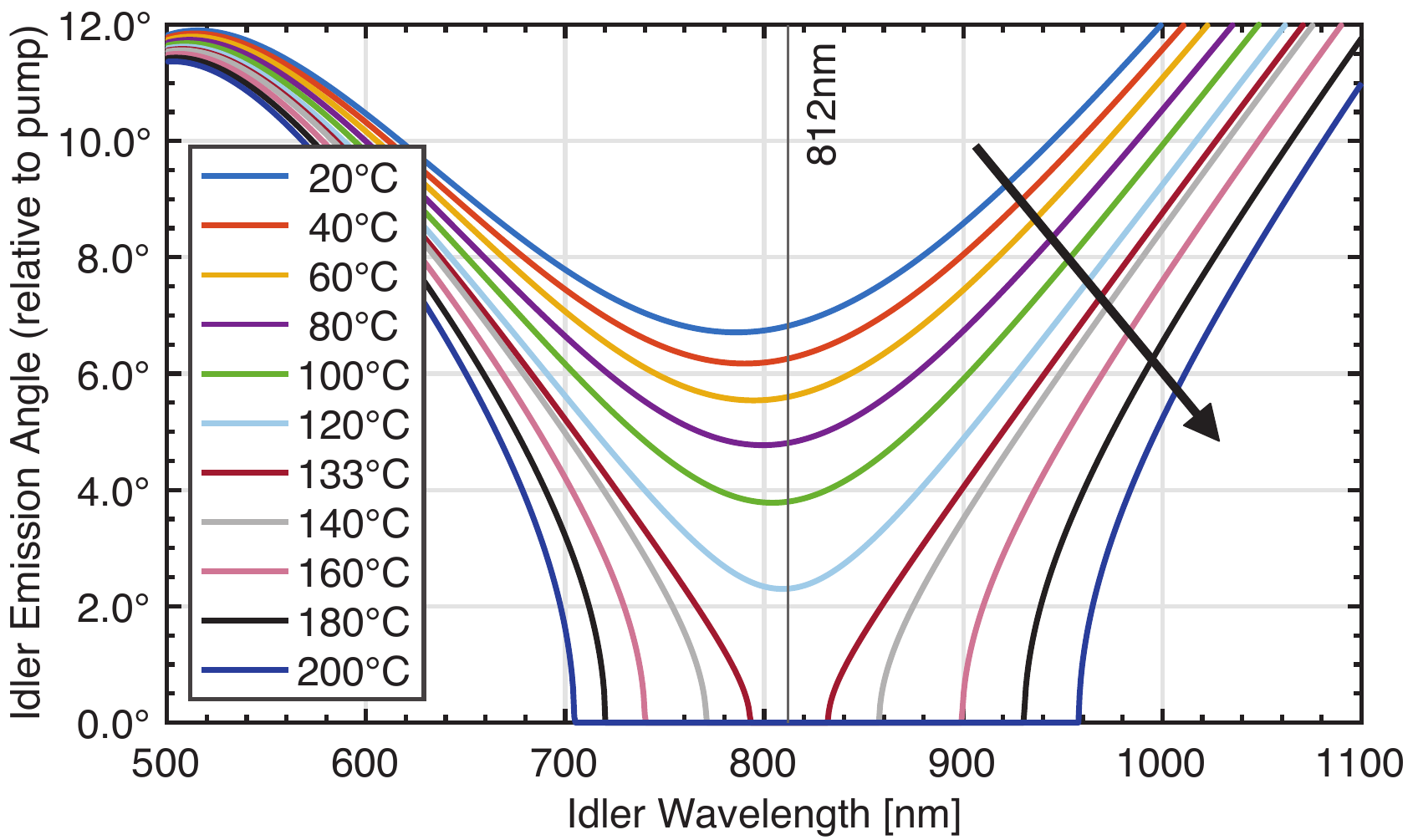}%
\caption{Broadband signal/idler emission angles in ppCLT with 9.5 $\mu$m poling period, showing how the SPDC source does not act as an ideal dipole emitter when broadband entangled photon pairs are desired. Points where curves reach a value of zero correspond to collinear, non-degenerate emission. Note that at certain temperatures, there need not be any SPDC emission in certain directions, governed by the spectral emission characteristics in Fig.\ref{ppCLT-spectrum-Temp}. Arrow indicating direction of increasing temperature.\label{coneAngles}}%
\end{figure}

\begin{figure}[ht]
\includegraphics[width=\columnwidth]{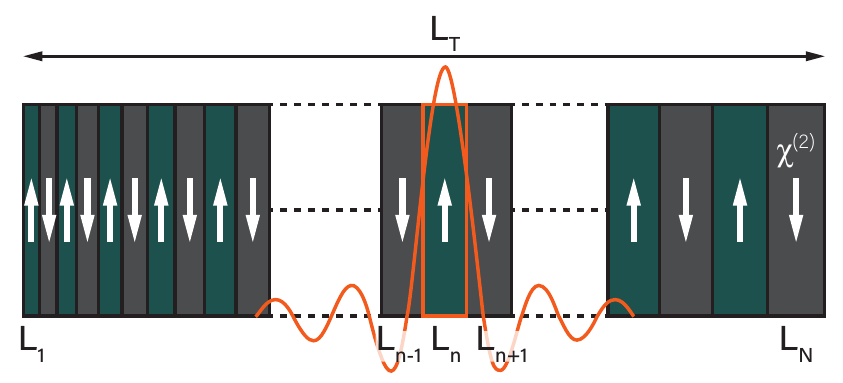}%
\caption{Depiction of a chirped periodically poled crystal with total length $L_T$. Each crystal segment of length $L_n$ is associated with a \textit{sinc}-shaped downconversion amplitude around a different center wavelength. Equivalently to a normal periodically poled crystal, the domain orientation is reversed in adjacent elements. \label{ppLT-crystal}}%
\end{figure}

Additionally, both signal and idler have an inherent angular emission property determined by the momentum conservation condition in Eq.\ref{deltaKQPM}. The strong frequency dependence of the outgoing angles can be numerically calculated from the momentum conservation equation if one decomposes the wave vectors into their longitudinal and transverse components, and solves the trigonometric equations for transverse phase-matching. Subsequently computing the emission angle for each frequency of interest, in conjunction with the frequency dependent refractive index of the nonlinear crystal, the SPDC cone's spectral characteristic for various crystal temperatures can be plotted (Fig.\ref{coneAngles}). As predicted, at the degenerate emission temperature of 133\degree C, the plot shows that the emission angle of the 812 nm signal and idler photons is 0\degree. More importantly however, comparing the results to those in Fig.\ref{ppCLT-spectrum-Temp}, the shaded non-degenerate region also corresponds to a collinear emission profile. Vice versa, temperatures below 133\degree C display a (albeit weak) degenerate, non-collinear character.
In a chirped crystal, the bandwidth of the down-converted entangled photons is increased by changing the poling period throughout the crystal. The increasing width elements correspond to different phase-matched wavelength configurations of Eq.\ref{deltaKQPM} (Fig.\ref{ppLT-crystal}). The photons still add up to the narrow linewidth of the pump laser but the temporal resolution is given by the temporal alignment and down-converted bandwidth of the entangled photons. This is why entangled photon experiments are said to have independent spectral and temporal resolutions. The importance of the pump laser's linewidth in determining the entangled photon state's purity, and thus the access to non-classical spectroscopic properties, is discussed in the SI.

\begin{figure}[ht]
\includegraphics[width=\columnwidth]{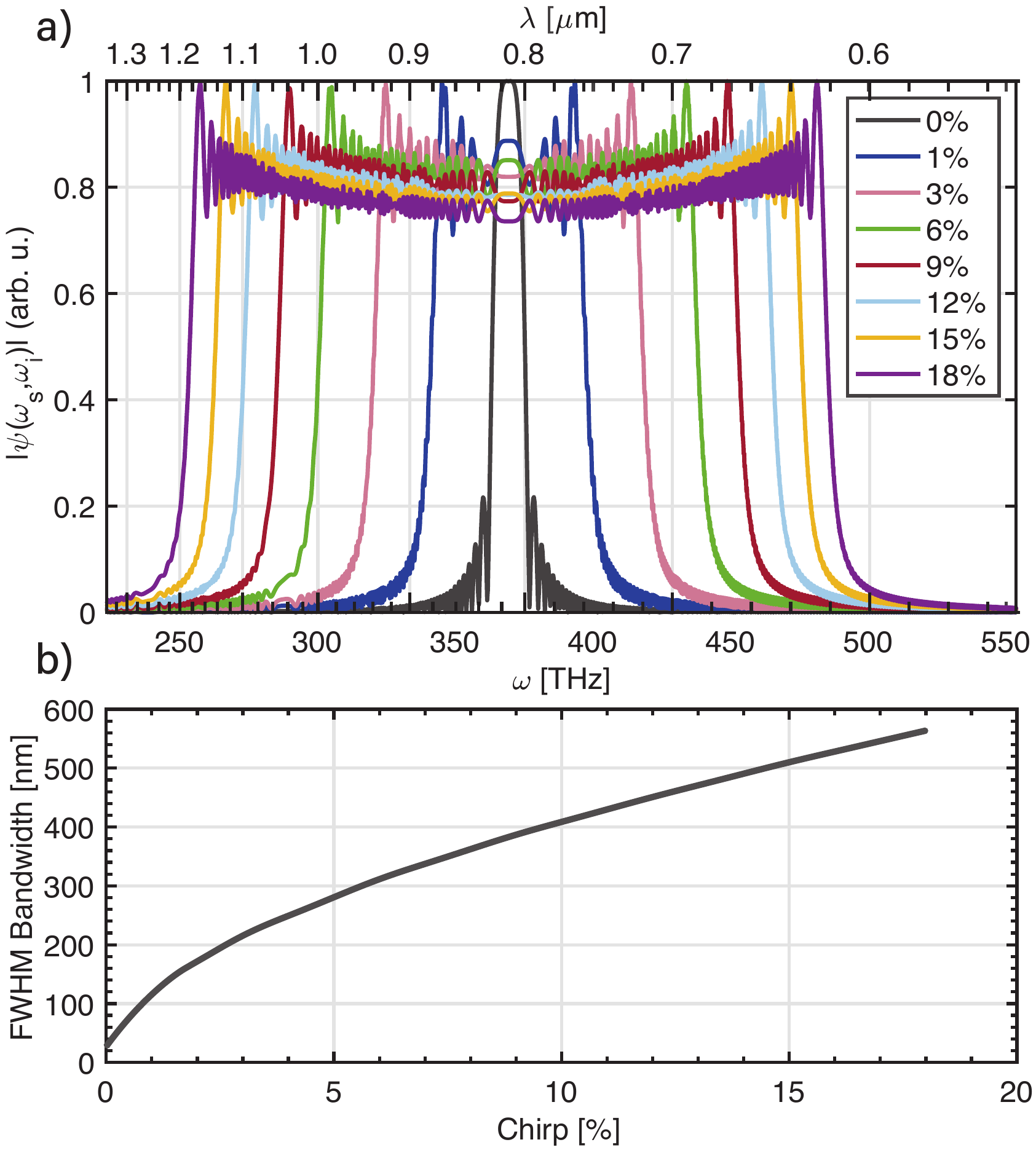}%
\caption{\textbf{a)} SPDC emission bandwidth broadening as a function of chirp parameter, defined as total percentage deviation from the degenerate poling period. \textbf{b)} Numerically calculated FWHM bandwidth trend as a function of poling period chirp. \label{chirp-sweep}}%
\end{figure}

Following a similar treatment to \cite{Atature2002}, the crystal can be decomposed into its individual domains, in each of which the downconversion amplitude follows the usual \textit{sinc} function-like phase-matching dependence \cite{Burlakov1997}. This is attributed to the Fourier transform of a rectangular function in real-space representing the probability of a downconversion event occurring inside the crystal element.
\begin{equation}
    A_n\left ( \omega_s,\omega_i \right )=L_nsinc\left ( \frac{L_n\Delta k}{2} \right )
\end{equation}
With $L_n$ being the length of the $n^{th}$ crystal element. Similarly, a cumulative-phase term can be defined, which is related to a forward propagating wave with phase-mismatch $\Delta k$ traversing a crystal length $L_{prop}$ after it has been generated in the SPDC process at crystal element \textit{n}. 
\begin{equation}
    \varphi = \Delta k L_{prop}=\Delta k\left ( \frac{L_n}{2}+\sum_{m=n+1}^{N}L_m \right )
\end{equation}
Carrying out a sum over all individual crystal elements then allows for the resulting SPDC spectrum to be calculated (Eq.\ref{chirpedSPDCeq}) and thereby observe the effects of different chirping parameters. For lithium tantalate, this is shown in Fig.\ref{chirp-sweep}. In the case of CLT, a $\pm$10\% chirp around the degenerate poling periodicity calculated above yields a full octave spanning SPDC bandwidth. Trivially, in the limit of no chirp, where all crystal elements are equal in size, the solution for a periodically poled crystal of length \textit{L} is recovered.
\begin{equation} \label{chirpedSPDCeq}
\begin{aligned}
    A\left ( \omega_s,\omega_i \right )=\chi_0\sum_{n=1}^{N}\left ( -1 \right )^nL_nsinc\left ( \frac{L_n\Delta k}{2} \right )\\
    \times e^{-i\Delta k\left ( \frac{L_n}{2}+\sum_{m=n+1}^{N}L_m \right )}
\end{aligned}
\end{equation}

\begin{figure*}[ht]
\includegraphics[width=\textwidth]{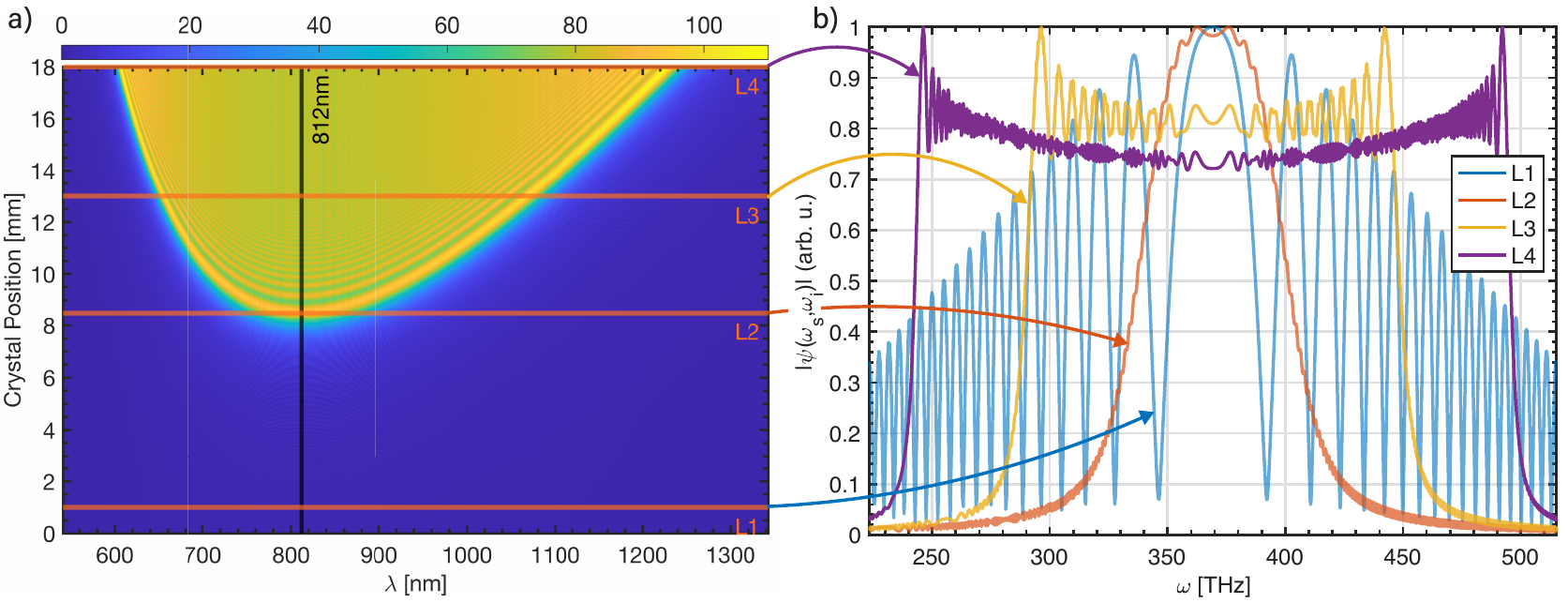}%
\caption{\textbf{a)} Build up of the collective SPDC spectrum as a function of the position inside a chirped ppCLT grating in the direction of propagation. \textbf{b)} Line cuts across the 2D spectrum plot in a) showing the normalized spectrum (in frequency space) at the positions designated as L1, L2, L3, and L4 respectively. \label{broadband-spectrum}}%
\end{figure*}

By plotting the solution of Eq.\ref{chirpedSPDCeq} as a function of the position inside the crystal along the propagation direction, we can gain some intuition for how the constructive interference results in a broadened spectrum. In the case of CLT, a $\pm$10\% chirp around the degenerate poling periodicity calculated above yields a full octave spanning SPDC bandwidth as shown in Fig.\ref{broadband-spectrum}. Here we used a total grating length of 18 mm equal to the chip used in subsequent experiments.

The 2D plot in Fig.\ref{broadband-spectrum} provides some additional insight into how modulation of the poling period affects the build-up of the emitted spectrum. In particular, it's important to note that while the first 8mm length of the crystal appears to show no downconversion taking place, this is indeed not the case. While all crystal elements in this region satisfy a particular phase-matching condition, it is only after a certain propagation length that the phases of each spectral mode coherently sum to a strong enough amplitude that would be perceivable on the plot. Furthermore, the broadening of the downconverted bandwidth does have a particular drawback, in that the frequencies close to degeneracy become weaker in intensity. This is best observed in the line cuts for positions L3 and L4, where instead of a constant intensity across all frequencies, the emission spectrum peaks at the lower and higher cut-offs but begins to droop more severely as the bandwidth increases. Thus, increasing the chirping of the poling period indefinitely does not necessarily yield a flat power spectral density. It should be pointed out that this issue can be addressed via chirp profiles which follow not a linear but some higher order functional form \cite{Branczyk2011}.

In contrast to narrowband SPDC where the emission angles for the photon pairs are well confined within the two narrow arcs satisfying momentum conservation, such extremely broadband downconversion poses inherent difficulties in the spatial degree of freedom due angular (chromatic) dispersion of the resulting beam. As Fig.\ref{coneAngles} shows, the emission angles increase quickly away from degeneracy. This presents several experimental issues, as the beam no longer represents an ideal point source, which will be discussed in the experimental section below. Taking into account the physical dimensions of the grating cross section (detailed in the experimental section below), it is easy to see that the output aperture of the grating can at most only accommodate $arctan(0.15mm/9mm)\approx1\degree$ of emission in the horizontal axis, and $arctan(0.5mm/9mm)\approx3.2\degree$ along the vertical axis relative to the center point of the crystal where we focus. Beyond this, large parts of the SPDC emission cone will begin to interact with the domain boundaries which form the grating channel. While in principle the refractive index is the same for both un-poled and poled crystal regions, the periodic poling process causes some stress at the domain walls. Therefore, the edge of the each pattern in the lateral direction acts as a kind of scattering center, possibly resulting in unwanted interference patterns. Therefore special attention should be paid to accommodate the spatial distribution of the cone as it diverges inside the channel.

A final point to consider is the total length of the grating.is to be chosen by considering two interlinked criteria. Firstly, the expected power spectral density of the entangled photons at the output scales as $L^2$. Emission intensity can therefore be significantly increased by utilizing longer structures. Conversely, the expected power spectral density also depends on a $sinc\left ( \frac{L_n\Delta k}{2} \right )$ multiplicative factor. Hence, longer gratings result in narrower emission bandwidths. This ‘dual’-argument also applies to poling patterns which have a chirp, as while increasing the amount of variation in the poling periodicity serves to broaden the emission bandwidth, each spectral component will be less populated with photons and thus less brilliant. 

\section{\label{experiment}Experiment}
\subsection{\label{spdc-characterization}SPDC Characterization}

The periodically poled congruent lithium tantalate gratings were manufactured by HC Photonics. The unchirped chip consists of 8 gratings with different poling periods (8.5, 9, 9.5, 10, 10.5, 11, 11.5, 12 \(\mu\)m). It is 20 mm long and each channel has a 0.9 mm by 0.5 mm cross section. The chirped chip consists of 6 gratings with varying chirp parameters and entrance poling periods (Table \ref{Chip-parameters}). Each grating is 18 mm long and has a 1 mm by 0.3 mm cross section. The gratings are AR-coated on both input and output faces at 406nm(R<0.5\%)/812nm(R<0.5\%).

\begin{table}[ht]
\centering
\begin{tabularx}{0.9\columnwidth} { 
  | >{\centering\arraybackslash}X
  | >{\centering\arraybackslash}X
  | >{\centering\arraybackslash}X
  | >{\centering\arraybackslash}X 
  | >{\centering\arraybackslash}X | }
  \hline
 Grating & $\Lambda_1$[$\mu$m] & $\Lambda_c$[$\mu$m] & $\Lambda_2$[$\mu$m] & Chirp \\ [0.5ex]
  \hline
  \hline
 1 & 9.000 & 9.45 & 9.900 & 10\% \\
 \hline
 2 & 9.048 & 9.50 & 9.952 & 10\% \\
 \hline
 3 & 9.095 & 9.55 & 10.005 & 10\% \\
 \hline
 4 & 9.143 & 9.60 & 10.057 & 10\% \\
 \hline
 5 & 9.190 & 9.65 & 10.110 & 10\% \\
 \hline
 6 & 9.238 & 9.70 & 10.162 & 10\% \\
 \hline
\end{tabularx}
\caption{Chip parameters}
\label{Chip-parameters}
\end{table}

The experimental setup, as illustrated in Fig.\ref{experimental-layout}, consisted of a computer-tunable, CW Ti:Sapphire laser (M2 SolsTiS) and an external SHG cavity to act as the pump source. The maximal output power of the SHG at 406 nm was 1.15 W with a 0.89 nm linewidth. The pump beam is focused through the ppCLT grating using an aspheric lens with a 40 cm focal length. This focal length is chosen such that the size of the focused beam was smaller than the grating's cross sectional dimension and further optimized for peak SPDC brightness. The position of the crystal can be finely adjusted in the XYZ directions using a piezoelectric stage, which is critical for maximizing the entangled photon flux. The angular deviation of the grating's long axis relative to the input beam was not as crucial as the X-Y centering of the chip around the focused beam and the z-control over exact focal point positioning. The temperature of the crystal is controlled by a ceramic heater and PID loop (Covesion) to an accuracy of 10 mK.

To measure the SPDC spectrum as well as the power of the downconverted flux, the output of the grating is passed through a total of three OD-4 500 nm longpass filters (Edmund Optics) to eliminate the residual 406 nm pump. The SPDC cone is collimated with an off-axis parabolic mirror and then focused into a USB spectrometer (Thorlabs). For more detailed characterization, the SPDC cone is collimated with a telephoto lens and then focused into a spectrometer (Princeton Instruments IsoPlane) with a 15cm focal length lens. In this configuration, spectral and power measurements are performed with an electron multiplying intensified CCD camera (emICCD, Princeton Instruments MAX4). For the broadband spectral measurements, spectra are collected and stitched together as a 800 nm blazed grating with 150 grooves/mm is scanned across several center wavelengths. These spectra are then background corrected, quantum efficiency corrected according to the detection efficiency of the emICCD camera, and normalized.

\begin{figure}[ht]
\includegraphics[width=\columnwidth]{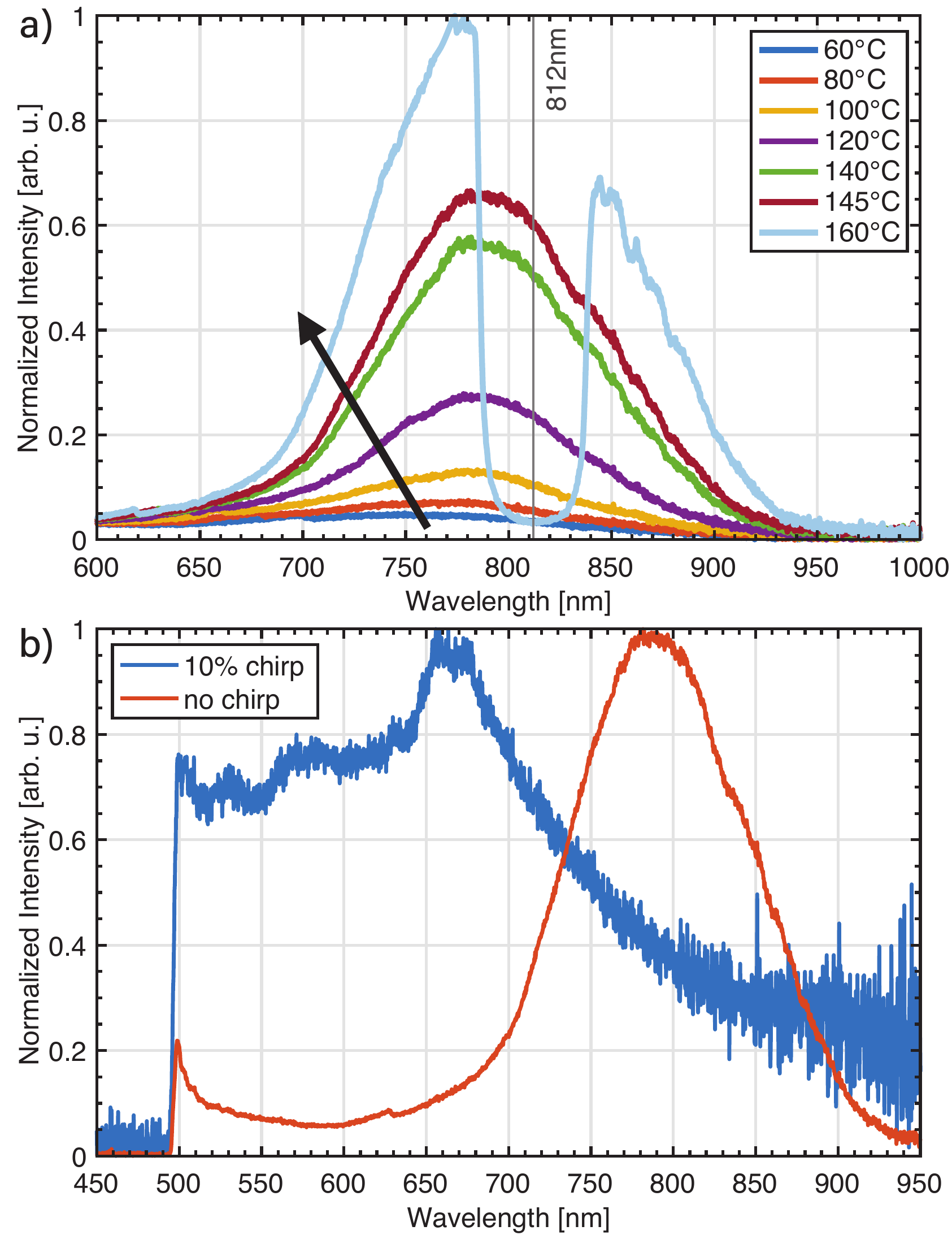}%
\caption{\textbf{a)} SPDC spectra for a 9.50 \textmu m unchirped grating, gray reference line marks the degenerate wavelength of 812 nm. Arrow a guide in the direction of increasing temperature. \textbf{b)} SPDC spectra for a chirped grating with a 9.50 \textmu m center poling period and 10\% chirp rate. Both spectra taken at a temperature of 145\degree C. The chirped emission is broader and has a flatter top than the unchirped emission. Note the detector cut-off at 900 nm.
\label{ppCLT-spectra}}%
\end{figure}

Fig.\ref{ppCLT-spectra} (a) depicts the SPDC spectra for the 9.50 \textmu m unchirped grating as the temperature of the crystal is varied from 60\degree C to 160\degree C. Above 150\degree C, the shift from degenerate to non-degenerate emission conditions is apparent from the decrease in intensity around the degenerate wavelength region of 812 nm (gray dashed line). In comparison, as shown in Fig.\ref{ppCLT-spectra} (b), at 145 \degree C, the chirped chip (blue curve) exhibits much broader and flatter emission than the unchirped counterpart (red curve).

\subsection{\label{two-photon-interference}Two-Photon Interference}

\begin{figure*}[ht]
\includegraphics[width=\textwidth]{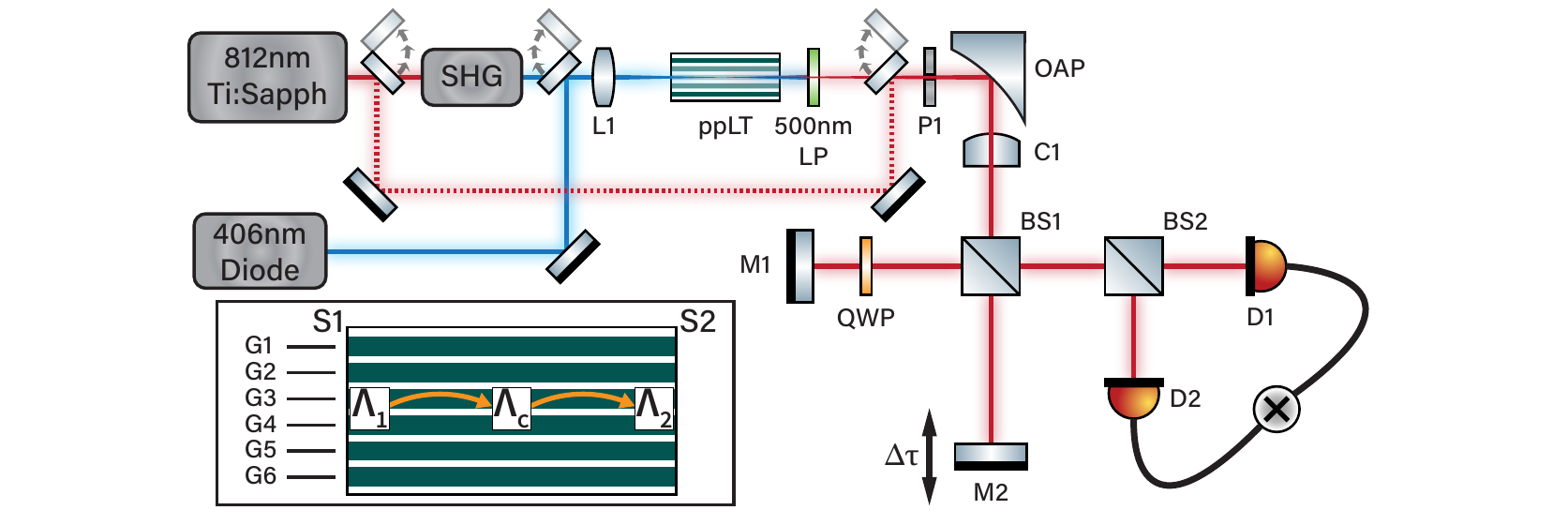}%
\caption{Entangled photon pair Michelson interferometer used to measure the fourth order interference. SHG: second harmonic generation unit, L1: focusing lens, ppLT: periodically poled lithium tantalate chip, LP: longpass filter, OAP: off-axis parabolic mirror, C1: cylindrical lens, BS1 and BS2: 50:50 beamsplitters, M1 and M2: mirrors, QWP: quarter-wave plate, D1 and D2: multimode fiber coupled single photon avalanche diodes connected to coincidence counting unit. The flip mirrors allow re-configuring such that the Ti:Sapph output is picked off and used for aid in alignment of downstream optics. A 406 nm diode laser is used as an alternate pump source. (Inset) Chip diagram showing the grating layout and their corresponding entrance ($\Lambda_1$), center ($\Lambda_c$), and end ($\Lambda_2$) poling periodicity locations in reference to Table.\ref{Chip-parameters}. \label{experimental-layout}}%
\end{figure*}

The fourth-order interference between entangled photon pairs was quantified with a two-photon Michelson interferometer, depicted in Fig.\ref{experimental-layout}. The Michelson interferometer is chosen because the configuration allows a broader bandwidth of the SPDC cone to be easily utilized, in contrast to a Mach-Zehnder configuration. Following the end of the ppLT chip, the entangled photons are collimated with an off-axis parabolic mirror and a cylindrical lens. These photons are directed through a broadband nonpolarizing 50:50 beamsplitter (BS1, Layertec) which separates the incident photons into two paths. In the reflected path, the photons pass through an achromatic quarter-waveplate (Thorlabs) twice to rotate their polarizations. The optical path length difference between the two arms is adjusted by scanning a mirror mounted to a linear stage in the transmitted arm. Coincidence counts vs. optical path length difference are then collected at one of the output arms of BS1. The coincidence counting setup consists of a second broadband 50:50 beamsplitter (BS2, Layertec), and two free-space-to-fiber coupling setups connecting to single-photon avalanche diodes (SPADs, Laser Component COUNT) and time-tagging electronics (PicoHarp 300). Because of the high SPDC flux, the pump beam must first be dimmed by six orders of magnitude to avoid SPAD saturation (counting rate $10^{6}$ photons/s). Full counting rate experiments can be completed using the CCD alone \cite{Reichert2018,Zhang2020} but were not implemented at the time of writing this paper.

The free-space-to-fiber coupling setups each consist of 2 mirrors to align the beamsplitter outputs through a 4.51 mm focal length asphere, which then focuses the beam into a multimode fiber (105 \(\mu m\) core diameter, 0.22 NA) connected to the SPADs. For these measurements, the coincidence time-bin resolution is set to 8 ps and coincidences are summed within a 10 ns window. Fourth-order interferograms were measured for different bandwidths of entangled photon pairs. For the first experiment, bandpass filters centered at 810 nm with a 10 nm FWHM are mounted to the front of each of the fiber coupling units. Note that this filter is not at the 812nm degenerate wavelength and reduces the degree of entanglement. For the second experiment, the 810 nm bandpass filters are removed to couple the broadband, collinear SPDC into the SPADs. Finally, the oven is allowed to cool to the non-collinear but degenerate temperature of 150\degree C, and broadband two-photon interference was measured again. Note that due to the large bandwidth of the SPDC, the fiber coupling efficiency of the non-degenerate photons is also poor at less than 10$\%$ of total power. As discussed at the end of this section, optics development is still needed similar to using classical broadband ultrafast lasers.

\begin{figure*}[ht]
\includegraphics[width=\textwidth]{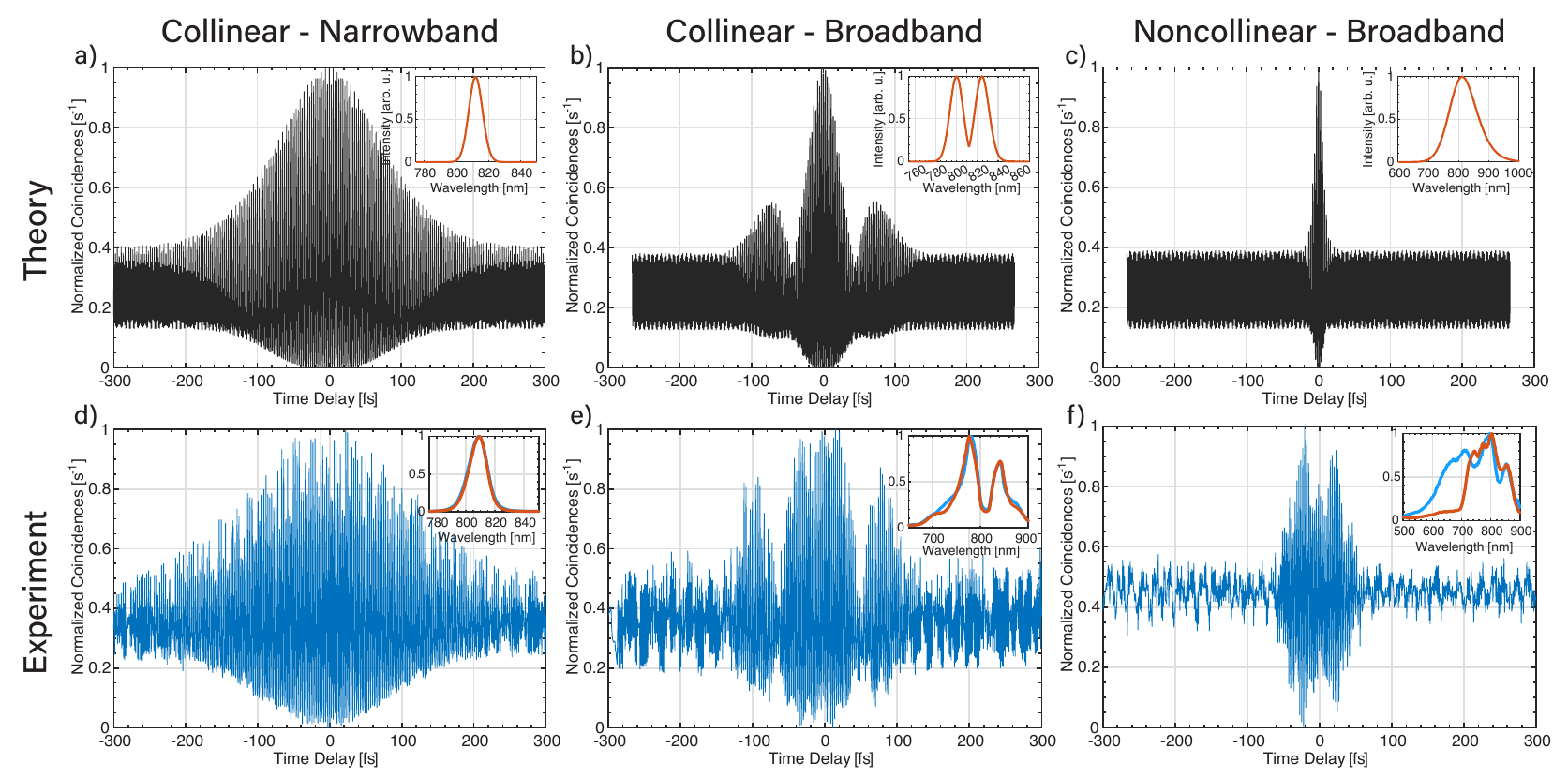}%
\caption{Simulated (top row) and measured (bottom row) fourth-order interference for narrowband collinear (a,d), broadband collinear (b,e), and broadband non-collinear (c,f) entangled photons. Coincidence counts are measured with 15 s integration time. The insets in each figure depict the SPDC spectra used to simulate or measure the corresponding interference in each detector arm. For simulations, Gaussian functions are used to approximate SPDC spectral envelopes. As shown in Table \ref{coherence-lengths}, the theroretical and measured coherence times are 200 fs and 245 fs for collinear narrowband, 45 fs and 57.4 fs for collinear broadband, 18 fs and 62 fs for non-collinear broadband. The growing difference between expected time resolution from the SPDC bandwidth and the measured interference is due to the difficulty with fiber coupling the full cone emission profile. \label{michelson-plots}}%
\end{figure*}

Fig.\ref{michelson-plots} shows the simulated fourth-order interferences using the produced bandwidth of the SPDC source and then the measured fourth-order interferences for two collinear bandwidths and a third non-collinear, full spectrum measurement. When the quarter-wave plate is set to 0\degree, the photons arriving at BS1 have the same polarization, and interfere coherently as correlated pairs. The interference pattern is robust against pump power fluctuations. When the quarter-wave plate is rotated to 45\degree, the reflected arm becomes perpendicularly polarized to the transmitted arm, the interference pattern disappears, and the coincidence counts follow the average pump power reading collected before and after the coincidence detection (SI Fig.8-9). The resulting coherence length of the total flux can be determined from the width of the interference peak. Table \ref{coherence-lengths} summarizes the theoretical and measured coherence lengths.

\begin{table}[ht]
\centering
\begin{tabularx}{0.9\columnwidth} { 
  | >{\centering\arraybackslash}X
  | >{\centering\arraybackslash}X
  | >{\centering\arraybackslash}X 
  | >{\centering\arraybackslash}X | }
  \hline
  & Collinear narrowband & Collinear broadband & Non-collinear broadband \\ [0.5ex]
  \hline
  \hline
 Theoretical coherence time [fs] & \hspace{2cm} 200 & \hspace{2cm} 45 & \hspace{2cm} 18 \\
 \hline
 Measured coherence time [fs] & \hspace{2cm} 245 & \hspace{2cm} 57.4 & \hspace{2cm} 62 \\
 \hline
\end{tabularx}
\caption{Coherence times of different configurations and bandwidths of SPDC}
\label{coherence-lengths}
\end{table}

For the measurement with 810-(10) nm bandpass filters mounted to the front of the fiber coupling units, shown in Fig.\ref{michelson-plots}(d), the coherence length is approximately 73 \textmu m (or 245 fs). The simulated interference pattern as well as the coherence length of 60\textmu m (200 fs), shown in Fig.\ref{michelson-plots}(a), agree fairly with the measurement given that the 10nm bandpass filter is offset from the degenerate wavelength of 812 nm. Note that the measured bandwidth after the fiber is roughly double what it actually is because the spectrometer slit had to be run wide to get sufficient flux for spectral analysis after the fiber coupling. The theoretical plot uses the correct bandwidth. Fig.\ref{michelson-plots}(e) shows the interference from collinear broadband SPDC flux without bandpass filters. The bandwidth is reduced and the SPDC spectrum is split (nondegenerate) because the phase matching temperature is maintained at the elevated collinear emission (Fig.\ref{ppCLT-spectrum-Temp}). From the width of the center peak, the coherence length is approximately 17.2 \textmu m (or 57.4 fs). For reference, the simulated coherence length is approximately 13 \textmu m (45 fs). Similarly, Fig.\ref{michelson-plots}(f) shows the lower temperature non-collinear broadband interference. Here, the full 200 nm bandwidth of the SPDC source is attempted to be used (expected temporal width 8fs). However the spectrum reaching the detector is limited to 125 nm because of collimation issues, upstream optics, and fiber coupling. The coherence length calculated from the measured width of the peak is 18.9 \textmu m (or 62 fs). This indicates that the degree of entanglement has degraded due to poor fiber coupling at non-collinear temperatures and other upstream optics. For example, the spectrum after the multiple beamsplitters is not balanced  (Fig.\ref{michelson-plots}(f), inset) as compared to the other cases. The increased temporal width is also indicative of the need for dispersion management upstream, just as is similar to a broadband ultrafast laser pulse of an expected ~20 fs pulse. The same would be true for the chirped (>200 nm) SPDC source so an interferometer is not shown here.

\begin{figure}[ht]
\includegraphics[width=\columnwidth]{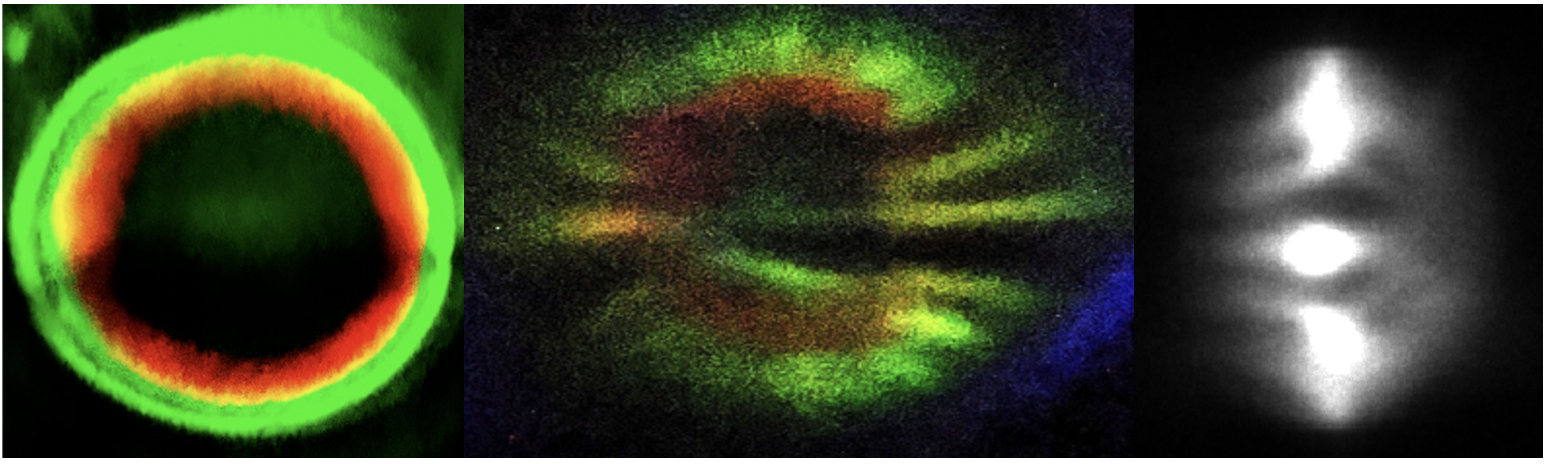}%
\caption{Images of collimated SPDC emission cone at 145 \degree C in the near field (left), far field (middle) after collimation with an off axis parabolic mirror, and far field after collimation and focusing by a 15.45cm focal length lens (right). The left and middle images are collected with a standard cell-phone camera, the right image is collected with an emICCD.
\label{beamimages}}%
\end{figure}

The lack of fiber coupling efficiency in the broadband, non-collinear case is not simply one of input optics. Immediately following the chip, the spatial profile of the entangled photons is cone-like with a wavelength-dependent distribution (Fig.\ref{beamimages} left). The measured angular deviation with wavelength matches the simulated entangled photon emission. Within a non-degenerate pair, the idler photon has a larger emission angle than the signal photon. The wavelength-dependent angular emission cone present a challenge in the implementation of free-space broadband entangled photon setups because the emission does not act like single point source. We attempted to collimate the emission cone using a variety of free-space optical configurations and found that a telephoto lens or an off-axis parabolic mirror (sometimes with an additional cylindrical lens for beam shaping) provides the best long range spatial profiles. After a propagation length of 45 cm into the far field, the entangled photon cone collapses (Fig.\ref{beamimages} middle). This spatial profile collapse is most likely due to the interference of photon pairs created along the length of the crystal and scattering effects near the domain boundaries of the grating. The end result shows characteristics of somewhere between a Laguerre and a Hermite mode which would match the circular emission in a square profile created by the source. The beam profile will likely be improved by using adaptive optics. The far field effects could also be reduced by using a waveguide instead of a grating; however, nanophotonic implementations cannot handle the high input powers used in this paper and fiber coupling of the broadband emission spectrum from the waveguide is not well developed. Another obvious option is, instead of using fiber coupled SPADs for the detection, is to move to emICCD photon counting techniques that would not require focusing of the collapsed beam profile. However, this approach still needs exploration and requires an emICCD even more costly than SPADs. Further investigation into the creation and collimation of broadband entangled photons is critical for the implementation of short temporal length, high flux entangled photon spectroscopy.

\section{\label{conclusion}Conclusion}
Quasi-phase-matching in periodically poled crystals offers significant benefits in the experimental implementation of broadband entangled photon sources for use in spectroscopy. Simulated parameters for unchirped, periodically poled lithium tantalate gratings were utilized to design and construct a broadband entangled photon source spanning nearly an octave in frequency. This broadband source was characterized by measuring the output spectra of the grating and the fourth order interference by a two-photon Michelson interferometer. The two-photon Michelson interferometer was used to emphasize where further optics and detector development is still need to utilize the broadband and short temporal width sources. As in, the SPDC spectrum can be reliably created from the designed chip, however, the down-stream optics and detectors still need development to fully utilize this source. Given the theoretical and experimentally demonstrated advantages of entangled photons in studies of chemical, biological, and material systems, we anticipate the widespread applicability of this particular energy-time entangled photon source.

\section*{\label{acknowledgment}Acknowledgment}
\noindent This material is based upon work supported by the U.S. Department of Energy, Office of Science, Office of Basic Energy Sciences, under Award Number DE-SC0020151.

\section*{Data Availability}
\noindent The data that support the findings of this study are available from the corresponding author upon reasonable request.

\bibliography{main.bib}

\providecommand{\noopsort}[1]{}\providecommand{\singleletter}[1]{#1}%
\begin{thebibliography}{53}%
\makeatletter
\providecommand \@ifxundefined [1]{%
 \@ifx{#1\undefined}
}%
\providecommand \@ifnum [1]{%
 \ifnum #1\expandafter \@firstoftwo
 \else \expandafter \@secondoftwo
 \fi
}%
\providecommand \@ifx [1]{%
 \ifx #1\expandafter \@firstoftwo
 \else \expandafter \@secondoftwo
 \fi
}%
\providecommand \natexlab [1]{#1}%
\providecommand \enquote  [1]{``#1''}%
\providecommand \bibnamefont  [1]{#1}%
\providecommand \bibfnamefont [1]{#1}%
\providecommand \citenamefont [1]{#1}%
\providecommand \href@noop [0]{\@secondoftwo}%
\providecommand \href [0]{\begingroup \@sanitize@url \@href}%
\providecommand \@href[1]{\@@startlink{#1}\@@href}%
\providecommand \@@href[1]{\endgroup#1\@@endlink}%
\providecommand \@sanitize@url [0]{\catcode `\\12\catcode `\$12\catcode
  `\&12\catcode `\#12\catcode `\^12\catcode `\_12\catcode `\%12\relax}%
\providecommand \@@startlink[1]{}%
\providecommand \@@endlink[0]{}%
\providecommand \url  [0]{\begingroup\@sanitize@url \@url }%
\providecommand \@url [1]{\endgroup\@href {#1}{\urlprefix }}%
\providecommand \urlprefix  [0]{URL }%
\providecommand \Eprint [0]{\href }%
\providecommand \doibase [0]{http://dx.doi.org/}%
\providecommand \selectlanguage [0]{\@gobble}%
\providecommand \bibinfo  [0]{\@secondoftwo}%
\providecommand \bibfield  [0]{\@secondoftwo}%
\providecommand \translation [1]{[#1]}%
\providecommand \BibitemOpen [0]{}%
\providecommand \bibitemStop [0]{}%
\providecommand \bibitemNoStop [0]{.\EOS\space}%
\providecommand \EOS [0]{\spacefactor3000\relax}%
\providecommand \BibitemShut  [1]{\csname bibitem#1\endcsname}%
\let\auto@bib@innerbib\@empty
\bibitem [{\citenamefont {Burnham}\ and\ \citenamefont
  {Weinberg}(1970)}]{Burnham1970}%
  \BibitemOpen
  \bibfield  {author} {\bibinfo {author} {\bibfnamefont {D.~C.}\ \bibnamefont
  {Burnham}}\ and\ \bibinfo {author} {\bibfnamefont {D.~L.}\ \bibnamefont
  {Weinberg}},\ }\bibfield  {title} {\enquote {\bibinfo {title} {Observation of
  simultaneity in parametric production of optical photon pairs},}\ }\href
  {\doibase 10.1103/PhysRevLett.25.84} {\bibfield  {journal} {\bibinfo
  {journal} {Phys. Rev. Lett.}\ }\textbf {\bibinfo {volume} {25}},\ \bibinfo
  {pages} {84--87} (\bibinfo {year} {1970})}\BibitemShut {NoStop}%
\bibitem [{\citenamefont {Szoke}\ \emph {et~al.}(2020)\citenamefont {Szoke},
  \citenamefont {Liu}, \citenamefont {Hickam}, \citenamefont {He},\ and\
  \citenamefont {Cushing}}]{Szoke2020}%
  \BibitemOpen
  \bibfield  {author} {\bibinfo {author} {\bibfnamefont {S.}~\bibnamefont
  {Szoke}}, \bibinfo {author} {\bibfnamefont {H.}~\bibnamefont {Liu}}, \bibinfo
  {author} {\bibfnamefont {B.~P.}\ \bibnamefont {Hickam}}, \bibinfo {author}
  {\bibfnamefont {M.}~\bibnamefont {He}}, \ and\ \bibinfo {author}
  {\bibfnamefont {S.~K.}\ \bibnamefont {Cushing}},\ }\bibfield  {title}
  {\enquote {\bibinfo {title} {Entangled light–matter interactions and
  spectroscopy},}\ }\href {\doibase 10.1039/D0TC02300K} {\bibfield  {journal}
  {\bibinfo  {journal} {J. Mater. Chem. C}\ }\textbf {\bibinfo {volume} {8}},\
  \bibinfo {pages} {10732--10741} (\bibinfo {year} {2020})}\BibitemShut
  {NoStop}%
\bibitem [{\citenamefont {Hong}, \citenamefont {Ou},\ and\ \citenamefont
  {Mandel}(1987)}]{Hong1987}%
  \BibitemOpen
  \bibfield  {author} {\bibinfo {author} {\bibfnamefont {C.~K.}\ \bibnamefont
  {Hong}}, \bibinfo {author} {\bibfnamefont {Z.~Y.}\ \bibnamefont {Ou}}, \ and\
  \bibinfo {author} {\bibfnamefont {L.}~\bibnamefont {Mandel}},\ }\bibfield
  {title} {\enquote {\bibinfo {title} {Measurement of subpicosecond time
  intervals between two photons by interference},}\ }\href {\doibase
  10.1103/PhysRevLett.59.2044} {\bibfield  {journal} {\bibinfo  {journal}
  {Phys. Rev. Lett.}\ }\textbf {\bibinfo {volume} {59}},\ \bibinfo {pages}
  {2044--2046} (\bibinfo {year} {1987})}\BibitemShut {NoStop}%
\bibitem [{\citenamefont {Mitev}\ \emph {et~al.}(2020)\citenamefont {Mitev},
  \citenamefont {Balet}, \citenamefont {Torcheboeuf}, \citenamefont {Renevey},\
  and\ \citenamefont {Boiko}}]{Mitev2020}%
  \BibitemOpen
  \bibfield  {author} {\bibinfo {author} {\bibfnamefont {V.}~\bibnamefont
  {Mitev}}, \bibinfo {author} {\bibfnamefont {L.}~\bibnamefont {Balet}},
  \bibinfo {author} {\bibfnamefont {N.}~\bibnamefont {Torcheboeuf}}, \bibinfo
  {author} {\bibfnamefont {P.}~\bibnamefont {Renevey}}, \ and\ \bibinfo
  {author} {\bibfnamefont {D.~L.}\ \bibnamefont {Boiko}},\ }\bibfield  {title}
  {\enquote {\bibinfo {title} {Discrimination of entangled photon pair from
  classical photons by de broglie wavelength},}\ }\href {\doibase
  10.1038/s41598-020-63833-8} {\bibfield  {journal} {\bibinfo  {journal}
  {Scientific Reports}\ }\textbf {\bibinfo {volume} {10}},\ \bibinfo {pages}
  {7087} (\bibinfo {year} {2020})}\BibitemShut {NoStop}%
\bibitem [{\citenamefont {Shimizu}, \citenamefont {Edamatsu},\ and\
  \citenamefont {Itoh}(2003)}]{Shimizu2003}%
  \BibitemOpen
  \bibfield  {author} {\bibinfo {author} {\bibfnamefont {R.}~\bibnamefont
  {Shimizu}}, \bibinfo {author} {\bibfnamefont {K.}~\bibnamefont {Edamatsu}}, \
  and\ \bibinfo {author} {\bibfnamefont {T.}~\bibnamefont {Itoh}},\ }\bibfield
  {title} {\enquote {\bibinfo {title} {Quantum diffraction and interference of
  spatially correlated photon pairs generated by spontaneous parametric
  down-conversion},}\ }\href {\doibase 10.1103/PhysRevA.67.041805} {\bibfield
  {journal} {\bibinfo  {journal} {Phys. Rev. A}\ }\textbf {\bibinfo {volume}
  {67}},\ \bibinfo {pages} {041805} (\bibinfo {year} {2003})}\BibitemShut
  {NoStop}%
\bibitem [{\citenamefont {Abouraddy}\ \emph {et~al.}(2001)\citenamefont
  {Abouraddy}, \citenamefont {Saleh}, \citenamefont {Sergienko},\ and\
  \citenamefont {Teich}}]{Abouraddy2001}%
  \BibitemOpen
  \bibfield  {author} {\bibinfo {author} {\bibfnamefont {A.~F.}\ \bibnamefont
  {Abouraddy}}, \bibinfo {author} {\bibfnamefont {B.~E.~A.}\ \bibnamefont
  {Saleh}}, \bibinfo {author} {\bibfnamefont {A.~V.}\ \bibnamefont
  {Sergienko}}, \ and\ \bibinfo {author} {\bibfnamefont {M.~C.}\ \bibnamefont
  {Teich}},\ }\bibfield  {title} {\enquote {\bibinfo {title} {Double-slit
  interference of biphotons generated in spontaneous parametric downconversion
  from a thick crystal},}\ }\href {\doibase 10.1088/1464-4266/3/1/359}
  {\bibfield  {journal} {\bibinfo  {journal} {Journal of Optics B: Quantum and
  Semiclassical Optics}\ }\textbf {\bibinfo {volume} {3}},\ \bibinfo {pages}
  {S50--S54} (\bibinfo {year} {2001})}\BibitemShut {NoStop}%
\bibitem [{\citenamefont {Javanainen}\ and\ \citenamefont
  {Gould}(1990)}]{Javanainen1990}%
  \BibitemOpen
  \bibfield  {author} {\bibinfo {author} {\bibfnamefont {J.}~\bibnamefont
  {Javanainen}}\ and\ \bibinfo {author} {\bibfnamefont {P.~L.}\ \bibnamefont
  {Gould}},\ }\bibfield  {title} {\enquote {\bibinfo {title} {Linear intensity
  dependence of a two-photon transition rate},}\ }\href {\doibase
  10.1103/PhysRevA.41.5088} {\bibfield  {journal} {\bibinfo  {journal} {Phys.
  Rev. A}\ }\textbf {\bibinfo {volume} {41}},\ \bibinfo {pages} {5088--5091}
  (\bibinfo {year} {1990})}\BibitemShut {NoStop}%
\bibitem [{\citenamefont {Dayan}\ \emph {et~al.}(2005)\citenamefont {Dayan},
  \citenamefont {Pe'er}, \citenamefont {Friesem},\ and\ \citenamefont
  {Silberberg}}]{Dayan2005}%
  \BibitemOpen
  \bibfield  {author} {\bibinfo {author} {\bibfnamefont {B.}~\bibnamefont
  {Dayan}}, \bibinfo {author} {\bibfnamefont {A.}~\bibnamefont {Pe'er}},
  \bibinfo {author} {\bibfnamefont {A.~A.}\ \bibnamefont {Friesem}}, \ and\
  \bibinfo {author} {\bibfnamefont {Y.}~\bibnamefont {Silberberg}},\ }\bibfield
   {title} {\enquote {\bibinfo {title} {Nonlinear interactions with an
  ultrahigh flux of broadband entangled photons},}\ }\href {\doibase
  10.1103/PhysRevLett.94.043602} {\bibfield  {journal} {\bibinfo  {journal}
  {Phys. Rev. Lett.}\ }\textbf {\bibinfo {volume} {94}},\ \bibinfo {pages}
  {043602} (\bibinfo {year} {2005})}\BibitemShut {NoStop}%
\bibitem [{\citenamefont {Villabona-Monsalve}, \citenamefont {Burdick},\ and\
  \citenamefont {Goodson}(2020)}]{Goodson2020}%
  \BibitemOpen
  \bibfield  {author} {\bibinfo {author} {\bibfnamefont {J.~P.}\ \bibnamefont
  {Villabona-Monsalve}}, \bibinfo {author} {\bibfnamefont {R.~K.}\ \bibnamefont
  {Burdick}}, \ and\ \bibinfo {author} {\bibfnamefont {T.}~\bibnamefont
  {Goodson}},\ }\bibfield  {title} {\enquote {\bibinfo {title} {Measurements of
  entangled two-photon absorption in organic molecules with cw-pumped type-i
  spontaneous parametric down-conversion},}\ }\href {\doibase
  10.1021/acs.jpcc.0c08678} {\bibfield  {journal} {\bibinfo  {journal} {The
  Journal of Physical Chemistry C}\ }\textbf {\bibinfo {volume} {124}},\
  \bibinfo {pages} {24526--24532} (\bibinfo {year} {2020})},\ \Eprint
  {http://arxiv.org/abs/https://doi.org/10.1021/acs.jpcc.0c08678}
  {https://doi.org/10.1021/acs.jpcc.0c08678} \BibitemShut {NoStop}%
\bibitem [{\citenamefont {Raymer}\ \emph {et~al.}(2013)\citenamefont {Raymer},
  \citenamefont {Marcus}, \citenamefont {Widom},\ and\ \citenamefont
  {Vitullo}}]{Raymer2013}%
  \BibitemOpen
  \bibfield  {author} {\bibinfo {author} {\bibfnamefont {M.~G.}\ \bibnamefont
  {Raymer}}, \bibinfo {author} {\bibfnamefont {A.~H.}\ \bibnamefont {Marcus}},
  \bibinfo {author} {\bibfnamefont {J.~R.}\ \bibnamefont {Widom}}, \ and\
  \bibinfo {author} {\bibfnamefont {D.~L.~P.}\ \bibnamefont {Vitullo}},\
  }\bibfield  {title} {\enquote {\bibinfo {title} {Entangled photon-pair
  two-dimensional fluorescence spectroscopy (epp-2dfs)},}\ }\href {\doibase
  10.1021/jp405829n} {\bibfield  {journal} {\bibinfo  {journal} {The Journal of
  Physical Chemistry B}\ }\textbf {\bibinfo {volume} {117}},\ \bibinfo {pages}
  {15559--15575} (\bibinfo {year} {2013})},\ \bibinfo {note} {pMID: 24047447},\
  \Eprint {http://arxiv.org/abs/https://doi.org/10.1021/jp405829n}
  {https://doi.org/10.1021/jp405829n} \BibitemShut {NoStop}%
\bibitem [{\citenamefont {Tabakaev}\ \emph {et~al.}(2021)\citenamefont
  {Tabakaev}, \citenamefont {Montagnese}, \citenamefont {Haack}, \citenamefont
  {Bonacina}, \citenamefont {Wolf}, \citenamefont {Zbinden},\ and\
  \citenamefont {Thew}}]{Tabakaev2021}%
  \BibitemOpen
  \bibfield  {author} {\bibinfo {author} {\bibfnamefont {D.}~\bibnamefont
  {Tabakaev}}, \bibinfo {author} {\bibfnamefont {M.}~\bibnamefont
  {Montagnese}}, \bibinfo {author} {\bibfnamefont {G.}~\bibnamefont {Haack}},
  \bibinfo {author} {\bibfnamefont {L.}~\bibnamefont {Bonacina}}, \bibinfo
  {author} {\bibfnamefont {J.-P.}\ \bibnamefont {Wolf}}, \bibinfo {author}
  {\bibfnamefont {H.}~\bibnamefont {Zbinden}}, \ and\ \bibinfo {author}
  {\bibfnamefont {R.~T.}\ \bibnamefont {Thew}},\ }\bibfield  {title} {\enquote
  {\bibinfo {title} {Energy-time-entangled two-photon molecular absorption},}\
  }\href {\doibase 10.1103/PhysRevA.103.033701} {\bibfield  {journal} {\bibinfo
   {journal} {Phys. Rev. A}\ }\textbf {\bibinfo {volume} {103}},\ \bibinfo
  {pages} {033701} (\bibinfo {year} {2021})}\BibitemShut {NoStop}%
\bibitem [{\citenamefont {Li}\ \emph {et~al.}(2020)\citenamefont {Li},
  \citenamefont {Li}, \citenamefont {Altuzarra}, \citenamefont {Classen},\ and\
  \citenamefont {Agarwal}}]{Li2020}%
  \BibitemOpen
  \bibfield  {author} {\bibinfo {author} {\bibfnamefont {T.}~\bibnamefont
  {Li}}, \bibinfo {author} {\bibfnamefont {F.}~\bibnamefont {Li}}, \bibinfo
  {author} {\bibfnamefont {C.}~\bibnamefont {Altuzarra}}, \bibinfo {author}
  {\bibfnamefont {A.}~\bibnamefont {Classen}}, \ and\ \bibinfo {author}
  {\bibfnamefont {G.~S.}\ \bibnamefont {Agarwal}},\ }\bibfield  {title}
  {\enquote {\bibinfo {title} {Squeezed light induced two-photon absorption
  fluorescence of fluorescein biomarkers},}\ }\href {\doibase
  10.1063/5.0010909} {\bibfield  {journal} {\bibinfo  {journal} {Applied
  Physics Letters}\ }\textbf {\bibinfo {volume} {116}},\ \bibinfo {pages}
  {254001} (\bibinfo {year} {2020})}\BibitemShut {NoStop}%
\bibitem [{\citenamefont {Kang}\ \emph {et~al.}(2020)\citenamefont {Kang},
  \citenamefont {Nasiri~Avanaki}, \citenamefont {Mosquera}, \citenamefont
  {Burdick}, \citenamefont {Villabona-Monsalve}, \citenamefont {Goodson},\ and\
  \citenamefont {Schatz}}]{Kang2020}%
  \BibitemOpen
  \bibfield  {author} {\bibinfo {author} {\bibfnamefont {G.}~\bibnamefont
  {Kang}}, \bibinfo {author} {\bibfnamefont {K.}~\bibnamefont
  {Nasiri~Avanaki}}, \bibinfo {author} {\bibfnamefont {M.~A.}\ \bibnamefont
  {Mosquera}}, \bibinfo {author} {\bibfnamefont {R.~K.}\ \bibnamefont
  {Burdick}}, \bibinfo {author} {\bibfnamefont {J.~P.}\ \bibnamefont
  {Villabona-Monsalve}}, \bibinfo {author} {\bibfnamefont {T.}~\bibnamefont
  {Goodson}}, \ and\ \bibinfo {author} {\bibfnamefont {G.~C.}\ \bibnamefont
  {Schatz}},\ }\bibfield  {title} {\enquote {\bibinfo {title} {Efficient
  modeling of organic chromophores for entangled two-photon absorption},}\
  }\href {\doibase 10.1021/jacs.0c02808} {\bibfield  {journal} {\bibinfo
  {journal} {Journal of the American Chemical Society}\ }\textbf {\bibinfo
  {volume} {142}},\ \bibinfo {pages} {10446--10458} (\bibinfo {year}
  {2020})}\BibitemShut {NoStop}%
\bibitem [{\citenamefont {Boto}\ \emph {et~al.}(2000)\citenamefont {Boto},
  \citenamefont {Kok}, \citenamefont {Abrams}, \citenamefont {Braunstein},
  \citenamefont {Williams},\ and\ \citenamefont {Dowling}}]{Boto2000}%
  \BibitemOpen
  \bibfield  {author} {\bibinfo {author} {\bibfnamefont {A.~N.}\ \bibnamefont
  {Boto}}, \bibinfo {author} {\bibfnamefont {P.}~\bibnamefont {Kok}}, \bibinfo
  {author} {\bibfnamefont {D.~S.}\ \bibnamefont {Abrams}}, \bibinfo {author}
  {\bibfnamefont {S.~L.}\ \bibnamefont {Braunstein}}, \bibinfo {author}
  {\bibfnamefont {C.~P.}\ \bibnamefont {Williams}}, \ and\ \bibinfo {author}
  {\bibfnamefont {J.~P.}\ \bibnamefont {Dowling}},\ }\bibfield  {title}
  {\enquote {\bibinfo {title} {Quantum interferometric optical lithography:
  Exploiting entanglement to beat the diffraction limit},}\ }\href@noop {}
  {\bibfield  {journal} {\bibinfo  {journal} {Phys. Rev. Lett.}\ }\textbf
  {\bibinfo {volume} {85}},\ \bibinfo {pages} {2733--2736} (\bibinfo {year}
  {2000})}\BibitemShut {NoStop}%
\bibitem [{\citenamefont {Steuernagel}(2004)}]{Steuernagel2004}%
  \BibitemOpen
  \bibfield  {author} {\bibinfo {author} {\bibfnamefont {O.}~\bibnamefont
  {Steuernagel}},\ }\bibfield  {title} {\enquote {\bibinfo {title} {On the
  concentration behaviour of entangled photons},}\ }\href {\doibase
  10.1088/1464-4266/6/6/021} {\bibfield  {journal} {\bibinfo  {journal} {J.
  Opt. B: Quantum Semiclass}\ }\textbf {\bibinfo {volume} {6}},\ \bibinfo
  {pages} {S606--S609} (\bibinfo {year} {2004})}\BibitemShut {NoStop}%
\bibitem [{\citenamefont {Oka}(2011)}]{Oka2011}%
  \BibitemOpen
  \bibfield  {author} {\bibinfo {author} {\bibfnamefont {H.}~\bibnamefont
  {Oka}},\ }\bibfield  {title} {\enquote {\bibinfo {title} {Selective
  two-photon excitation of a vibronic state by correlated photons},}\ }\href
  {\doibase 10.1063/1.3573565} {\bibfield  {journal} {\bibinfo  {journal} {The
  Journal of Chemical Physics}\ }\textbf {\bibinfo {volume} {134}},\ \bibinfo
  {pages} {124313} (\bibinfo {year} {2011})},\ \Eprint
  {http://arxiv.org/abs/https://doi.org/10.1063/1.3573565}
  {https://doi.org/10.1063/1.3573565} \BibitemShut {NoStop}%
\bibitem [{\citenamefont {Oka}(2018)}]{Oka2018}%
  \BibitemOpen
  \bibfield  {author} {\bibinfo {author} {\bibfnamefont {H.}~\bibnamefont
  {Oka}},\ }\bibfield  {title} {\enquote {\bibinfo {title} {Enhanced
  vibrational-mode-selective two-step excitation using ultrabroadband
  frequency-entangled photons},}\ }\href {\doibase 10.1103/PhysRevA.97.063859}
  {\bibfield  {journal} {\bibinfo  {journal} {Phys. Rev. A}\ }\textbf {\bibinfo
  {volume} {97}},\ \bibinfo {pages} {063859} (\bibinfo {year}
  {2018})}\BibitemShut {NoStop}%
\bibitem [{\citenamefont {Schlawin}, \citenamefont {Dorfman},\ and\
  \citenamefont {Mukamel}(2018)}]{Schlawin2018}%
  \BibitemOpen
  \bibfield  {author} {\bibinfo {author} {\bibfnamefont {F.}~\bibnamefont
  {Schlawin}}, \bibinfo {author} {\bibfnamefont {K.~E.}\ \bibnamefont
  {Dorfman}}, \ and\ \bibinfo {author} {\bibfnamefont {S.}~\bibnamefont
  {Mukamel}},\ }\bibfield  {title} {\enquote {\bibinfo {title} {Entangled
  two-photon absorption spectroscopy},}\ }\href {\doibase
  10.1021/acs.accounts.8b00173} {\bibfield  {journal} {\bibinfo  {journal}
  {Accounts of Chemical Research}\ }\textbf {\bibinfo {volume} {51}},\ \bibinfo
  {pages} {2207--2214} (\bibinfo {year} {2018})},\ \bibinfo {note} {pMID:
  30179458},\ \Eprint
  {http://arxiv.org/abs/https://doi.org/10.1021/acs.accounts.8b00173}
  {https://doi.org/10.1021/acs.accounts.8b00173} \BibitemShut {NoStop}%
\bibitem [{\citenamefont {Dorfman}, \citenamefont {Schlawin},\ and\
  \citenamefont {Mukamel}(2016)}]{Dorfman2016}%
  \BibitemOpen
  \bibfield  {author} {\bibinfo {author} {\bibfnamefont {K.~E.}\ \bibnamefont
  {Dorfman}}, \bibinfo {author} {\bibfnamefont {F.}~\bibnamefont {Schlawin}}, \
  and\ \bibinfo {author} {\bibfnamefont {S.}~\bibnamefont {Mukamel}},\
  }\bibfield  {title} {\enquote {\bibinfo {title} {Nonlinear optical signals
  and spectroscopy with quantum light},}\ }\href {\doibase
  10.1103/RevModPhys.88.045008} {\bibfield  {journal} {\bibinfo  {journal}
  {Rev. Mod. Phys.}\ }\textbf {\bibinfo {volume} {88}},\ \bibinfo {pages}
  {045008} (\bibinfo {year} {2016})}\BibitemShut {NoStop}%
\bibitem [{\citenamefont {MacLean}, \citenamefont {Schwarz},\ and\
  \citenamefont {Resch}(2019)}]{Maclean2019}%
  \BibitemOpen
  \bibfield  {author} {\bibinfo {author} {\bibfnamefont {J.-P.~W.}\
  \bibnamefont {MacLean}}, \bibinfo {author} {\bibfnamefont {S.}~\bibnamefont
  {Schwarz}}, \ and\ \bibinfo {author} {\bibfnamefont {K.~J.}\ \bibnamefont
  {Resch}},\ }\bibfield  {title} {\enquote {\bibinfo {title} {Reconstructing
  ultrafast energy-time-entangled two-photon pulses},}\ }\href {\doibase
  10.1103/PhysRevA.100.033834} {\bibfield  {journal} {\bibinfo  {journal}
  {Phys. Rev. A}\ }\textbf {\bibinfo {volume} {100}},\ \bibinfo {pages}
  {033834} (\bibinfo {year} {2019})}\BibitemShut {NoStop}%
\bibitem [{\citenamefont {Nasr}\ \emph {et~al.}(2005)\citenamefont {Nasr},
  \citenamefont {Giuseppe}, \citenamefont {Saleh}, \citenamefont {Sergienko},\
  and\ \citenamefont {Teich}}]{Magued2005}%
  \BibitemOpen
  \bibfield  {author} {\bibinfo {author} {\bibfnamefont {M.~B.}\ \bibnamefont
  {Nasr}}, \bibinfo {author} {\bibfnamefont {G.~D.}\ \bibnamefont {Giuseppe}},
  \bibinfo {author} {\bibfnamefont {B.~E.}\ \bibnamefont {Saleh}}, \bibinfo
  {author} {\bibfnamefont {A.~V.}\ \bibnamefont {Sergienko}}, \ and\ \bibinfo
  {author} {\bibfnamefont {M.~C.}\ \bibnamefont {Teich}},\ }\bibfield  {title}
  {\enquote {\bibinfo {title} {Generation of high-flux ultra-broadband light by
  bandwidth amplification in spontaneous parametric down conversion},}\ }\href
  {\doibase https://doi.org/10.1016/j.optcom.2004.11.008} {\bibfield  {journal}
  {\bibinfo  {journal} {Optics Communications}\ }\textbf {\bibinfo {volume}
  {246}},\ \bibinfo {pages} {521--528} (\bibinfo {year} {2005})}\BibitemShut
  {NoStop}%
\bibitem [{\citenamefont {Kwiat}\ \emph {et~al.}(1995)\citenamefont {Kwiat},
  \citenamefont {Mattle}, \citenamefont {Weinfurter}, \citenamefont
  {Zeilinger}, \citenamefont {Sergienko},\ and\ \citenamefont
  {Shih}}]{Kwiat1995}%
  \BibitemOpen
  \bibfield  {author} {\bibinfo {author} {\bibfnamefont {P.~G.}\ \bibnamefont
  {Kwiat}}, \bibinfo {author} {\bibfnamefont {K.}~\bibnamefont {Mattle}},
  \bibinfo {author} {\bibfnamefont {H.}~\bibnamefont {Weinfurter}}, \bibinfo
  {author} {\bibfnamefont {A.}~\bibnamefont {Zeilinger}}, \bibinfo {author}
  {\bibfnamefont {A.~V.}\ \bibnamefont {Sergienko}}, \ and\ \bibinfo {author}
  {\bibfnamefont {Y.}~\bibnamefont {Shih}},\ }\bibfield  {title} {\enquote
  {\bibinfo {title} {New high-intensity source of polarization-entangled photon
  pairs},}\ }\href {\doibase 10.1103/PhysRevLett.75.4337} {\bibfield  {journal}
  {\bibinfo  {journal} {Phys. Rev. Lett.}\ }\textbf {\bibinfo {volume} {75}},\
  \bibinfo {pages} {4337--4341} (\bibinfo {year} {1995})}\BibitemShut {NoStop}%
\bibitem [{\citenamefont {Friberg}, \citenamefont {Hong},\ and\ \citenamefont
  {Mandel}(1985)}]{Friberg1985}%
  \BibitemOpen
  \bibfield  {author} {\bibinfo {author} {\bibfnamefont {S.}~\bibnamefont
  {Friberg}}, \bibinfo {author} {\bibfnamefont {C.~K.}\ \bibnamefont {Hong}}, \
  and\ \bibinfo {author} {\bibfnamefont {L.}~\bibnamefont {Mandel}},\
  }\bibfield  {title} {\enquote {\bibinfo {title} {Measurement of time delays
  in the parametric production of photon pairs},}\ }\href {\doibase
  10.1103/PhysRevLett.54.2011} {\bibfield  {journal} {\bibinfo  {journal}
  {Phys. Rev. Lett.}\ }\textbf {\bibinfo {volume} {54}},\ \bibinfo {pages}
  {2011--2013} (\bibinfo {year} {1985})}\BibitemShut {NoStop}%
\bibitem [{\citenamefont {Kwiat}\ \emph {et~al.}(1999)\citenamefont {Kwiat},
  \citenamefont {Waks}, \citenamefont {White}, \citenamefont {Appelbaum},\ and\
  \citenamefont {Eberhard}}]{Kwiat1999}%
  \BibitemOpen
  \bibfield  {author} {\bibinfo {author} {\bibfnamefont {P.~G.}\ \bibnamefont
  {Kwiat}}, \bibinfo {author} {\bibfnamefont {E.}~\bibnamefont {Waks}},
  \bibinfo {author} {\bibfnamefont {A.~G.}\ \bibnamefont {White}}, \bibinfo
  {author} {\bibfnamefont {I.}~\bibnamefont {Appelbaum}}, \ and\ \bibinfo
  {author} {\bibfnamefont {P.~H.}\ \bibnamefont {Eberhard}},\ }\bibfield
  {title} {\enquote {\bibinfo {title} {Ultrabright source of
  polarization-entangled photons},}\ }\href {\doibase 10.1103/PhysRevA.60.R773}
  {\bibfield  {journal} {\bibinfo  {journal} {Phys. Rev. A}\ }\textbf {\bibinfo
  {volume} {60}},\ \bibinfo {pages} {R773--R776} (\bibinfo {year}
  {1999})}\BibitemShut {NoStop}%
\bibitem [{\citenamefont {Karan}\ \emph {et~al.}(2020)\citenamefont {Karan},
  \citenamefont {Aarav}, \citenamefont {Bharadhwaj}, \citenamefont {Taneja},
  \citenamefont {De}, \citenamefont {Kulkarni}, \citenamefont {Meher},\ and\
  \citenamefont {Jha}}]{Karan2020}%
  \BibitemOpen
  \bibfield  {author} {\bibinfo {author} {\bibfnamefont {S.}~\bibnamefont
  {Karan}}, \bibinfo {author} {\bibfnamefont {S.}~\bibnamefont {Aarav}},
  \bibinfo {author} {\bibfnamefont {H.}~\bibnamefont {Bharadhwaj}}, \bibinfo
  {author} {\bibfnamefont {L.}~\bibnamefont {Taneja}}, \bibinfo {author}
  {\bibfnamefont {A.}~\bibnamefont {De}}, \bibinfo {author} {\bibfnamefont
  {G.}~\bibnamefont {Kulkarni}}, \bibinfo {author} {\bibfnamefont
  {N.}~\bibnamefont {Meher}}, \ and\ \bibinfo {author} {\bibfnamefont {A.~K.}\
  \bibnamefont {Jha}},\ }\bibfield  {title} {\enquote {\bibinfo {title} {Phase
  matching in beta-barium borate crystals for spontaneous parametric
  down-conversion},}\ }\href {\doibase 10.1088/2040-8986/ab89e4} {\bibfield
  {journal} {\bibinfo  {journal} {Journal of Optics}\ }\textbf {\bibinfo
  {volume} {22}},\ \bibinfo {pages} {083501} (\bibinfo {year}
  {2020})}\BibitemShut {NoStop}%
\bibitem [{\citenamefont {Nikogosyan}(1991)}]{Nikogosyan1991}%
  \BibitemOpen
  \bibfield  {author} {\bibinfo {author} {\bibfnamefont {D.~N.}\ \bibnamefont
  {Nikogosyan}},\ }\bibfield  {title} {\enquote {\bibinfo {title} {Beta barium
  borate (bbo)},}\ }\href {\doibase 10.1007/BF00323647} {\bibfield  {journal}
  {\bibinfo  {journal} {Applied Physics A}\ }\textbf {\bibinfo {volume} {52}},\
  \bibinfo {pages} {359--368} (\bibinfo {year} {1991})}\BibitemShut {NoStop}%
\bibitem [{\citenamefont {Midwinter}\ and\ \citenamefont
  {Warner}(1965)}]{Midwinter1965}%
  \BibitemOpen
  \bibfield  {author} {\bibinfo {author} {\bibfnamefont {J.~E.}\ \bibnamefont
  {Midwinter}}\ and\ \bibinfo {author} {\bibfnamefont {J.}~\bibnamefont
  {Warner}},\ }\bibfield  {title} {\enquote {\bibinfo {title} {The effects of
  phase matching method and of crystal symmetry on the polar dependence of
  third-order non-linear optical polarization},}\ }\href {\doibase
  10.1088/0508-3443/16/11/307} {\bibfield  {journal} {\bibinfo  {journal}
  {British Journal of Applied Physics}\ }\textbf {\bibinfo {volume} {16}},\
  \bibinfo {pages} {1667--1674} (\bibinfo {year} {1965})}\BibitemShut {NoStop}%
\bibitem [{\citenamefont {Roslyak}\ and\ \citenamefont
  {Mukamel}(2009)}]{Roslyak2009}%
  \BibitemOpen
  \bibfield  {author} {\bibinfo {author} {\bibfnamefont {O.}~\bibnamefont
  {Roslyak}}\ and\ \bibinfo {author} {\bibfnamefont {S.}~\bibnamefont
  {Mukamel}},\ }\bibfield  {title} {\enquote {\bibinfo {title}
  {Multidimensional pump-probe spectroscopy with entangled twin-photon
  states},}\ }\href {\doibase 10.1103/PhysRevA.79.063409} {\bibfield  {journal}
  {\bibinfo  {journal} {Phys. Rev. A}\ }\textbf {\bibinfo {volume} {79}},\
  \bibinfo {pages} {063409} (\bibinfo {year} {2009})}\BibitemShut {NoStop}%
\bibitem [{\citenamefont {Wang}\ \emph {et~al.}(1999)\citenamefont {Wang},
  \citenamefont {Pasiskevicius}, \citenamefont {Hellstr\"{o}m}, \citenamefont
  {Laurell},\ and\ \citenamefont {Karlsson}}]{Wang1999}%
  \BibitemOpen
  \bibfield  {author} {\bibinfo {author} {\bibfnamefont {S.}~\bibnamefont
  {Wang}}, \bibinfo {author} {\bibfnamefont {V.}~\bibnamefont {Pasiskevicius}},
  \bibinfo {author} {\bibfnamefont {J.}~\bibnamefont {Hellstr\"{o}m}}, \bibinfo
  {author} {\bibfnamefont {F.}~\bibnamefont {Laurell}}, \ and\ \bibinfo
  {author} {\bibfnamefont {H.}~\bibnamefont {Karlsson}},\ }\bibfield  {title}
  {\enquote {\bibinfo {title} {First-order type ii quasi-phase-matched uv
  generation in periodically poled ktp},}\ }\href {\doibase
  10.1364/OL.24.000978} {\bibfield  {journal} {\bibinfo  {journal} {Opt.
  Lett.}\ }\textbf {\bibinfo {volume} {24}},\ \bibinfo {pages} {978--980}
  (\bibinfo {year} {1999})}\BibitemShut {NoStop}%
\bibitem [{\citenamefont {Yu}\ \emph {et~al.}(2002)\citenamefont {Yu},
  \citenamefont {Ro}, \citenamefont {Cha}, \citenamefont {Kurimura},\ and\
  \citenamefont {Taira}}]{Yu2002}%
  \BibitemOpen
  \bibfield  {author} {\bibinfo {author} {\bibfnamefont {N.~E.}\ \bibnamefont
  {Yu}}, \bibinfo {author} {\bibfnamefont {J.~H.}\ \bibnamefont {Ro}}, \bibinfo
  {author} {\bibfnamefont {M.}~\bibnamefont {Cha}}, \bibinfo {author}
  {\bibfnamefont {S.}~\bibnamefont {Kurimura}}, \ and\ \bibinfo {author}
  {\bibfnamefont {T.}~\bibnamefont {Taira}},\ }\bibfield  {title} {\enquote
  {\bibinfo {title} {Broadband quasi-phase-matched second-harmonic generation
  in mgo-doped periodically poled linbo3 at the communications band},}\ }\href
  {\doibase 10.1364/OL.27.001046} {\bibfield  {journal} {\bibinfo  {journal}
  {Opt. Lett.}\ }\textbf {\bibinfo {volume} {27}},\ \bibinfo {pages}
  {1046--1048} (\bibinfo {year} {2002})}\BibitemShut {NoStop}%
\bibitem [{\citenamefont {Lin}\ \emph {et~al.}(2019)\citenamefont {Lin},
  \citenamefont {Yao}, \citenamefont {Hao}, \citenamefont {Zhang},
  \citenamefont {Mao}, \citenamefont {Wang}, \citenamefont {Chu}, \citenamefont
  {Wu}, \citenamefont {Fang}, \citenamefont {Qiao}, \citenamefont {Fang},
  \citenamefont {Bo},\ and\ \citenamefont {Cheng}}]{Lin2019}%
  \BibitemOpen
  \bibfield  {author} {\bibinfo {author} {\bibfnamefont {J.}~\bibnamefont
  {Lin}}, \bibinfo {author} {\bibfnamefont {N.}~\bibnamefont {Yao}}, \bibinfo
  {author} {\bibfnamefont {Z.}~\bibnamefont {Hao}}, \bibinfo {author}
  {\bibfnamefont {J.}~\bibnamefont {Zhang}}, \bibinfo {author} {\bibfnamefont
  {W.}~\bibnamefont {Mao}}, \bibinfo {author} {\bibfnamefont {M.}~\bibnamefont
  {Wang}}, \bibinfo {author} {\bibfnamefont {W.}~\bibnamefont {Chu}}, \bibinfo
  {author} {\bibfnamefont {R.}~\bibnamefont {Wu}}, \bibinfo {author}
  {\bibfnamefont {Z.}~\bibnamefont {Fang}}, \bibinfo {author} {\bibfnamefont
  {L.}~\bibnamefont {Qiao}}, \bibinfo {author} {\bibfnamefont {W.}~\bibnamefont
  {Fang}}, \bibinfo {author} {\bibfnamefont {F.}~\bibnamefont {Bo}}, \ and\
  \bibinfo {author} {\bibfnamefont {Y.}~\bibnamefont {Cheng}},\ }\bibfield
  {title} {\enquote {\bibinfo {title} {Broadband quasi-phase-matched harmonic
  generation in an on-chip monocrystalline lithium niobate microdisk
  resonator},}\ }\href {\doibase 10.1103/PhysRevLett.122.173903} {\bibfield
  {journal} {\bibinfo  {journal} {Phys. Rev. Lett.}\ }\textbf {\bibinfo
  {volume} {122}},\ \bibinfo {pages} {173903} (\bibinfo {year}
  {2019})}\BibitemShut {NoStop}%
\bibitem [{\citenamefont {Chen}\ \emph {et~al.}(2009)\citenamefont {Chen},
  \citenamefont {Pearlman}, \citenamefont {Ling}, \citenamefont {Fan},\ and\
  \citenamefont {Migdall}}]{Chen2009}%
  \BibitemOpen
  \bibfield  {author} {\bibinfo {author} {\bibfnamefont {J.}~\bibnamefont
  {Chen}}, \bibinfo {author} {\bibfnamefont {A.~J.}\ \bibnamefont {Pearlman}},
  \bibinfo {author} {\bibfnamefont {A.}~\bibnamefont {Ling}}, \bibinfo {author}
  {\bibfnamefont {J.}~\bibnamefont {Fan}}, \ and\ \bibinfo {author}
  {\bibfnamefont {A.}~\bibnamefont {Migdall}},\ }\bibfield  {title} {\enquote
  {\bibinfo {title} {A versatile waveguide source of photon pairs for
  chip-scale quantum information processing},}\ }\href {\doibase
  10.1364/OE.17.006727} {\bibfield  {journal} {\bibinfo  {journal} {Opt.
  Express}\ }\textbf {\bibinfo {volume} {17}},\ \bibinfo {pages} {6727--6740}
  (\bibinfo {year} {2009})}\BibitemShut {NoStop}%
\bibitem [{\citenamefont {Bock}\ \emph {et~al.}(2016)\citenamefont {Bock},
  \citenamefont {Lenhard}, \citenamefont {Chunnilall},\ and\ \citenamefont
  {Becher}}]{Bock2016}%
  \BibitemOpen
  \bibfield  {author} {\bibinfo {author} {\bibfnamefont {M.}~\bibnamefont
  {Bock}}, \bibinfo {author} {\bibfnamefont {A.}~\bibnamefont {Lenhard}},
  \bibinfo {author} {\bibfnamefont {C.}~\bibnamefont {Chunnilall}}, \ and\
  \bibinfo {author} {\bibfnamefont {C.}~\bibnamefont {Becher}},\ }\bibfield
  {title} {\enquote {\bibinfo {title} {Highly efficient heralded single-photon
  source for telecom wavelengths based on a ppln waveguide},}\ }\href {\doibase
  10.1364/OE.24.023992} {\bibfield  {journal} {\bibinfo  {journal} {Opt.
  Express}\ }\textbf {\bibinfo {volume} {24}},\ \bibinfo {pages} {23992--24001}
  (\bibinfo {year} {2016})}\BibitemShut {NoStop}%
\bibitem [{\citenamefont {Shi}\ and\ \citenamefont {Tomita}(2004)}]{Shi2004}%
  \BibitemOpen
  \bibfield  {author} {\bibinfo {author} {\bibfnamefont {B.-S.}\ \bibnamefont
  {Shi}}\ and\ \bibinfo {author} {\bibfnamefont {A.}~\bibnamefont {Tomita}},\
  }\bibfield  {title} {\enquote {\bibinfo {title} {Highly efficient generation
  of pulsed photon pairs with bulk periodically poled potassium titanyl
  phosphate},}\ }\href {\doibase 10.1364/JOSAB.21.002081} {\bibfield  {journal}
  {\bibinfo  {journal} {J. Opt. Soc. Am. B}\ }\textbf {\bibinfo {volume}
  {21}},\ \bibinfo {pages} {2081--2084} (\bibinfo {year} {2004})}\BibitemShut
  {NoStop}%
\bibitem [{\citenamefont {Armstrong}\ \emph {et~al.}(1962)\citenamefont
  {Armstrong}, \citenamefont {Bloembergen}, \citenamefont {Ducuing},\ and\
  \citenamefont {Pershan}}]{Armstrong1962}%
  \BibitemOpen
  \bibfield  {author} {\bibinfo {author} {\bibfnamefont {J.~A.}\ \bibnamefont
  {Armstrong}}, \bibinfo {author} {\bibfnamefont {N.}~\bibnamefont
  {Bloembergen}}, \bibinfo {author} {\bibfnamefont {J.}~\bibnamefont
  {Ducuing}}, \ and\ \bibinfo {author} {\bibfnamefont {P.~S.}\ \bibnamefont
  {Pershan}},\ }\bibfield  {title} {\enquote {\bibinfo {title} {Interactions
  between light waves in a nonlinear dielectric},}\ }\href {\doibase
  10.1103/PhysRev.127.1918} {\bibfield  {journal} {\bibinfo  {journal} {Phys.
  Rev.}\ }\textbf {\bibinfo {volume} {127}},\ \bibinfo {pages} {1918--1939}
  (\bibinfo {year} {1962})}\BibitemShut {NoStop}%
\bibitem [{\citenamefont {Tanzilli}\ \emph {et~al.}(2001)\citenamefont
  {Tanzilli}, \citenamefont {Riedmatten}, \citenamefont {Tittel}, \citenamefont
  {Zbinden}, \citenamefont {Baldi}, \citenamefont {Micheli}, \citenamefont
  {Ostrowsky},\ and\ \citenamefont {Gisin}}]{Tanzilli2001}%
  \BibitemOpen
  \bibfield  {author} {\bibinfo {author} {\bibfnamefont {S.}~\bibnamefont
  {Tanzilli}}, \bibinfo {author} {\bibfnamefont {H.~D.}\ \bibnamefont
  {Riedmatten}}, \bibinfo {author} {\bibfnamefont {W.}~\bibnamefont {Tittel}},
  \bibinfo {author} {\bibfnamefont {H.}~\bibnamefont {Zbinden}}, \bibinfo
  {author} {\bibfnamefont {P.}~\bibnamefont {Baldi}}, \bibinfo {author}
  {\bibfnamefont {M.~D.}\ \bibnamefont {Micheli}}, \bibinfo {author}
  {\bibfnamefont {D.}~\bibnamefont {Ostrowsky}}, \ and\ \bibinfo {author}
  {\bibfnamefont {N.}~\bibnamefont {Gisin}},\ }\bibfield  {title} {\enquote
  {\bibinfo {title} {Highly efficient photon-pair source using periodically
  poled lithium niobate waveguide},}\ }\href
  {https://digital-library.theiet.org/content/journals/10.1049/el_20010009}
  {\bibfield  {journal} {\bibinfo  {journal} {Electronics Letters}\ }\textbf
  {\bibinfo {volume} {37}},\ \bibinfo {pages} {26--28} (\bibinfo {year}
  {2001})}\BibitemShut {NoStop}%
\bibitem [{\citenamefont {Nasr}\ \emph {et~al.}(2008)\citenamefont {Nasr},
  \citenamefont {Carrasco}, \citenamefont {Saleh}, \citenamefont {Sergienko},
  \citenamefont {Teich}, \citenamefont {Torres}, \citenamefont {Torner},
  \citenamefont {Hum},\ and\ \citenamefont {Fejer}}]{Nasr2008}%
  \BibitemOpen
  \bibfield  {author} {\bibinfo {author} {\bibfnamefont {M.~B.}\ \bibnamefont
  {Nasr}}, \bibinfo {author} {\bibfnamefont {S.}~\bibnamefont {Carrasco}},
  \bibinfo {author} {\bibfnamefont {B.~E.~A.}\ \bibnamefont {Saleh}}, \bibinfo
  {author} {\bibfnamefont {A.~V.}\ \bibnamefont {Sergienko}}, \bibinfo {author}
  {\bibfnamefont {M.~C.}\ \bibnamefont {Teich}}, \bibinfo {author}
  {\bibfnamefont {J.~P.}\ \bibnamefont {Torres}}, \bibinfo {author}
  {\bibfnamefont {L.}~\bibnamefont {Torner}}, \bibinfo {author} {\bibfnamefont
  {D.~S.}\ \bibnamefont {Hum}}, \ and\ \bibinfo {author} {\bibfnamefont
  {M.~M.}\ \bibnamefont {Fejer}},\ }\bibfield  {title} {\enquote {\bibinfo
  {title} {Ultrabroadband biphotons generated via chirped quasi-phase-matched
  optical parametric down-conversion},}\ }\href {\doibase
  10.1103/PhysRevLett.100.183601} {\bibfield  {journal} {\bibinfo  {journal}
  {Phys. Rev. Lett.}\ }\textbf {\bibinfo {volume} {100}},\ \bibinfo {pages}
  {183601} (\bibinfo {year} {2008})}\BibitemShut {NoStop}%
\bibitem [{\citenamefont {Pe'er}\ \emph {et~al.}(2005)\citenamefont {Pe'er},
  \citenamefont {Dayan}, \citenamefont {Friesem},\ and\ \citenamefont
  {Silberberg}}]{Peer2005}%
  \BibitemOpen
  \bibfield  {author} {\bibinfo {author} {\bibfnamefont {A.}~\bibnamefont
  {Pe'er}}, \bibinfo {author} {\bibfnamefont {B.}~\bibnamefont {Dayan}},
  \bibinfo {author} {\bibfnamefont {A.~A.}\ \bibnamefont {Friesem}}, \ and\
  \bibinfo {author} {\bibfnamefont {Y.}~\bibnamefont {Silberberg}},\ }\bibfield
   {title} {\enquote {\bibinfo {title} {Temporal shaping of entangled
  photons},}\ }\href {\doibase 10.1103/PhysRevLett.94.073601} {\bibfield
  {journal} {\bibinfo  {journal} {Phys. Rev. Lett.}\ }\textbf {\bibinfo
  {volume} {94}},\ \bibinfo {pages} {073601} (\bibinfo {year}
  {2005})}\BibitemShut {NoStop}%
\bibitem [{\citenamefont {Ou}\ and\ \citenamefont {Lu}(1999)}]{Ou1999}%
  \BibitemOpen
  \bibfield  {author} {\bibinfo {author} {\bibfnamefont {Z.~Y.}\ \bibnamefont
  {Ou}}\ and\ \bibinfo {author} {\bibfnamefont {Y.~J.}\ \bibnamefont {Lu}},\
  }\bibfield  {title} {\enquote {\bibinfo {title} {Cavity enhanced spontaneous
  parametric down-conversion for the prolongation of correlation time between
  conjugate photons},}\ }\href {\doibase 10.1103/PhysRevLett.83.2556}
  {\bibfield  {journal} {\bibinfo  {journal} {Phys. Rev. Lett.}\ }\textbf
  {\bibinfo {volume} {83}},\ \bibinfo {pages} {2556--2559} (\bibinfo {year}
  {1999})}\BibitemShut {NoStop}%
\bibitem [{\citenamefont {Hadfield}(2009)}]{Hadfield2009}%
  \BibitemOpen
  \bibfield  {author} {\bibinfo {author} {\bibfnamefont {R.~H.}\ \bibnamefont
  {Hadfield}},\ }\bibfield  {title} {\enquote {\bibinfo {title} {Single-photon
  detectors for optical quantum information applications},}\ }\href {\doibase
  10.1038/nphoton.2009.230} {\bibfield  {journal} {\bibinfo  {journal} {Nature
  Photonics}\ }\textbf {\bibinfo {volume} {3}},\ \bibinfo {pages} {696--705}
  (\bibinfo {year} {2009})}\BibitemShut {NoStop}%
\bibitem [{\citenamefont {Zverev}\ \emph {et~al.}(1972)\citenamefont {Zverev},
  \citenamefont {Levchuk}, \citenamefont {Pashkov},\ and\ \citenamefont
  {Poryadin}}]{Zverev1972}%
  \BibitemOpen
  \bibfield  {author} {\bibinfo {author} {\bibfnamefont {G.~M.}\ \bibnamefont
  {Zverev}}, \bibinfo {author} {\bibfnamefont {E.~A.}\ \bibnamefont {Levchuk}},
  \bibinfo {author} {\bibfnamefont {V.~A.}\ \bibnamefont {Pashkov}}, \ and\
  \bibinfo {author} {\bibfnamefont {Y.~D.}\ \bibnamefont {Poryadin}},\
  }\bibfield  {title} {\enquote {\bibinfo {title} {Laser-radiation-induced
  damage to the surface of lithium niobate and tantalate single crystals},}\
  }\href {\doibase 10.1070/qe1972v002n02abeh004409} {\bibfield  {journal}
  {\bibinfo  {journal} {Soviet Journal of Quantum Electronics}\ }\textbf
  {\bibinfo {volume} {2}},\ \bibinfo {pages} {167--169} (\bibinfo {year}
  {1972})}\BibitemShut {NoStop}%
\bibitem [{\citenamefont {Antonov}\ \emph {et~al.}(1975)\citenamefont
  {Antonov}, \citenamefont {Arsenev}, \citenamefont {Linda},\ and\
  \citenamefont {Farstendiker}}]{Antonov1975}%
  \BibitemOpen
  \bibfield  {author} {\bibinfo {author} {\bibfnamefont {V.~A.}\ \bibnamefont
  {Antonov}}, \bibinfo {author} {\bibfnamefont {P.~A.}\ \bibnamefont
  {Arsenev}}, \bibinfo {author} {\bibfnamefont {I.~G.}\ \bibnamefont {Linda}},
  \ and\ \bibinfo {author} {\bibfnamefont {V.~L.}\ \bibnamefont
  {Farstendiker}},\ }\bibfield  {title} {\enquote {\bibinfo {title} {Colour
  centres in single crystals of lithium tantalate},}\ }\href {\doibase
  https://doi.org/10.1002/pssa.2210280234} {\bibfield  {journal} {\bibinfo
  {journal} {physica status solidi (a)}\ }\textbf {\bibinfo {volume} {28}},\
  \bibinfo {pages} {673--676} (\bibinfo {year} {1975})},\ \Eprint
  {http://arxiv.org/abs/https://onlinelibrary.wiley.com/doi/pdf/10.1002/pssa.2210280234}
  {https://onlinelibrary.wiley.com/doi/pdf/10.1002/pssa.2210280234}
  \BibitemShut {NoStop}%
\bibitem [{\citenamefont {Meyn}\ and\ \citenamefont {Fejer}(1997)}]{Meyn1997}%
  \BibitemOpen
  \bibfield  {author} {\bibinfo {author} {\bibfnamefont {J.-P.}\ \bibnamefont
  {Meyn}}\ and\ \bibinfo {author} {\bibfnamefont {M.~M.}\ \bibnamefont
  {Fejer}},\ }\bibfield  {title} {\enquote {\bibinfo {title} {Tunable
  ultraviolet radiation by second-harmonic generation in periodically poled
  lithium tantalate},}\ }\href {\doibase 10.1364/OL.22.001214} {\bibfield
  {journal} {\bibinfo  {journal} {Opt. Lett.}\ }\textbf {\bibinfo {volume}
  {22}},\ \bibinfo {pages} {1214--1216} (\bibinfo {year} {1997})}\BibitemShut
  {NoStop}%
\bibitem [{\citenamefont {Lopez-Mago}\ and\ \citenamefont
  {Novotny}(2012)}]{LopezMago2012}%
  \BibitemOpen
  \bibfield  {author} {\bibinfo {author} {\bibfnamefont {D.}~\bibnamefont
  {Lopez-Mago}}\ and\ \bibinfo {author} {\bibfnamefont {L.}~\bibnamefont
  {Novotny}},\ }\bibfield  {title} {\enquote {\bibinfo {title} {Coherence
  measurements with the two-photon michelson interferometer},}\ }\href
  {\doibase 10.1103/PhysRevA.86.023820} {\bibfield  {journal} {\bibinfo
  {journal} {Phys. Rev. A}\ }\textbf {\bibinfo {volume} {86}},\ \bibinfo
  {pages} {023820} (\bibinfo {year} {2012})}\BibitemShut {NoStop}%
\bibitem [{\citenamefont {Klyshko}, \citenamefont {Penin},\ and\ \citenamefont
  {Polkovnikov}(1970)}]{Klyshko1970}%
  \BibitemOpen
  \bibfield  {author} {\bibinfo {author} {\bibfnamefont {D.~N.}\ \bibnamefont
  {Klyshko}}, \bibinfo {author} {\bibfnamefont {A.~N.}\ \bibnamefont {Penin}},
  \ and\ \bibinfo {author} {\bibfnamefont {B.~F.}\ \bibnamefont
  {Polkovnikov}},\ }\bibfield  {title} {\enquote {\bibinfo {title} {Parametric
  luminescence and light scattering by polaritons},}\ }\href
  {http://www.jetpletters.ac.ru/ps/1714/article_26042.shtml} {\bibfield
  {journal} {\bibinfo  {journal} {JETP Lett.}\ }\textbf {\bibinfo {volume}
  {11}},\ \bibinfo {pages} {11--14} (\bibinfo {year} {1970})}\BibitemShut
  {NoStop}%
\bibitem [{\citenamefont {Boyd}\ and\ \citenamefont {Prato}(2008)}]{Boyd2008}%
  \BibitemOpen
  \bibfield  {author} {\bibinfo {author} {\bibfnamefont {R.}~\bibnamefont
  {Boyd}}\ and\ \bibinfo {author} {\bibfnamefont {D.}~\bibnamefont {Prato}},\
  }\href {https://books.google.com/books?id=uoRUi1Yb7ooC} {\emph {\bibinfo
  {title} {Nonlinear Optics}}},\ \bibinfo {edition} {3rd}\ ed.\ (\bibinfo
  {publisher} {Elsevier Science},\ \bibinfo {year} {2008})\BibitemShut
  {NoStop}%
\bibitem [{\citenamefont {{Fejer}}\ \emph {et~al.}(1992)\citenamefont
  {{Fejer}}, \citenamefont {{Magel}}, \citenamefont {{Jundt}},\ and\
  \citenamefont {{Byer}}}]{fejer1992}%
  \BibitemOpen
  \bibfield  {author} {\bibinfo {author} {\bibfnamefont {M.~M.}\ \bibnamefont
  {{Fejer}}}, \bibinfo {author} {\bibfnamefont {G.~A.}\ \bibnamefont
  {{Magel}}}, \bibinfo {author} {\bibfnamefont {D.~H.}\ \bibnamefont
  {{Jundt}}}, \ and\ \bibinfo {author} {\bibfnamefont {R.~L.}\ \bibnamefont
  {{Byer}}},\ }\bibfield  {title} {\enquote {\bibinfo {title}
  {Quasi-phase-matched second harmonic generation: tuning and tolerances},}\
  }\href {\doibase 10.1109/3.161322} {\bibfield  {journal} {\bibinfo  {journal}
  {IEEE Journal of Quantum Electronics}\ }\textbf {\bibinfo {volume} {28}},\
  \bibinfo {pages} {2631--2654} (\bibinfo {year} {1992})}\BibitemShut {NoStop}%
\bibitem [{\citenamefont {Moutzouris}\ \emph {et~al.}(2011)\citenamefont
  {Moutzouris}, \citenamefont {Hloupis}, \citenamefont {Stavrakas},
  \citenamefont {Triantis},\ and\ \citenamefont {Chou}}]{Moutzouris2011}%
  \BibitemOpen
  \bibfield  {author} {\bibinfo {author} {\bibfnamefont {K.}~\bibnamefont
  {Moutzouris}}, \bibinfo {author} {\bibfnamefont {G.}~\bibnamefont {Hloupis}},
  \bibinfo {author} {\bibfnamefont {I.}~\bibnamefont {Stavrakas}}, \bibinfo
  {author} {\bibfnamefont {D.}~\bibnamefont {Triantis}}, \ and\ \bibinfo
  {author} {\bibfnamefont {M.-H.}\ \bibnamefont {Chou}},\ }\bibfield  {title}
  {\enquote {\bibinfo {title} {Temperature-dependent visible to near-infrared
  optical properties of 8 mol\% mg-doped lithium tantalate},}\ }\href {\doibase
  10.1364/OME.1.000458} {\bibfield  {journal} {\bibinfo  {journal} {Opt. Mater.
  Express}\ }\textbf {\bibinfo {volume} {1}},\ \bibinfo {pages} {458--465}
  (\bibinfo {year} {2011})}\BibitemShut {NoStop}%
\bibitem [{\citenamefont {Di~Giuseppe}\ \emph {et~al.}(2002)\citenamefont
  {Di~Giuseppe}, \citenamefont {Atat\"ure}, \citenamefont {Shaw}, \citenamefont
  {Sergienko}, \citenamefont {Saleh},\ and\ \citenamefont
  {Teich}}]{Atature2002}%
  \BibitemOpen
  \bibfield  {author} {\bibinfo {author} {\bibfnamefont {G.}~\bibnamefont
  {Di~Giuseppe}}, \bibinfo {author} {\bibfnamefont {M.}~\bibnamefont
  {Atat\"ure}}, \bibinfo {author} {\bibfnamefont {M.~D.}\ \bibnamefont {Shaw}},
  \bibinfo {author} {\bibfnamefont {A.~V.}\ \bibnamefont {Sergienko}}, \bibinfo
  {author} {\bibfnamefont {B.~E.~A.}\ \bibnamefont {Saleh}}, \ and\ \bibinfo
  {author} {\bibfnamefont {M.~C.}\ \bibnamefont {Teich}},\ }\bibfield  {title}
  {\enquote {\bibinfo {title} {Entangled-photon generation from parametric
  down-conversion in media with inhomogeneous nonlinearity},}\ }\href {\doibase
  10.1103/PhysRevA.66.013801} {\bibfield  {journal} {\bibinfo  {journal} {Phys.
  Rev. A}\ }\textbf {\bibinfo {volume} {66}},\ \bibinfo {pages} {013801}
  (\bibinfo {year} {2002})}\BibitemShut {NoStop}%
\bibitem [{\citenamefont {Burlakov}\ \emph {et~al.}(1997)\citenamefont
  {Burlakov}, \citenamefont {Chekhova}, \citenamefont {Klyshko}, \citenamefont
  {Kulik}, \citenamefont {Penin}, \citenamefont {Shih},\ and\ \citenamefont
  {Strekalov}}]{Burlakov1997}%
  \BibitemOpen
  \bibfield  {author} {\bibinfo {author} {\bibfnamefont {A.~V.}\ \bibnamefont
  {Burlakov}}, \bibinfo {author} {\bibfnamefont {M.~V.}\ \bibnamefont
  {Chekhova}}, \bibinfo {author} {\bibfnamefont {D.~N.}\ \bibnamefont
  {Klyshko}}, \bibinfo {author} {\bibfnamefont {S.~P.}\ \bibnamefont {Kulik}},
  \bibinfo {author} {\bibfnamefont {A.~N.}\ \bibnamefont {Penin}}, \bibinfo
  {author} {\bibfnamefont {Y.~H.}\ \bibnamefont {Shih}}, \ and\ \bibinfo
  {author} {\bibfnamefont {D.~V.}\ \bibnamefont {Strekalov}},\ }\bibfield
  {title} {\enquote {\bibinfo {title} {Interference effects in spontaneous
  two-photon parametric scattering from two macroscopic regions},}\ }\href
  {\doibase 10.1103/PhysRevA.56.3214} {\bibfield  {journal} {\bibinfo
  {journal} {Phys. Rev. A}\ }\textbf {\bibinfo {volume} {56}},\ \bibinfo
  {pages} {3214--3225} (\bibinfo {year} {1997})}\BibitemShut {NoStop}%
\bibitem [{\citenamefont {Bra\'{n}czyk}\ \emph {et~al.}(2011)\citenamefont
  {Bra\'{n}czyk}, \citenamefont {Fedrizzi}, \citenamefont {Stace},
  \citenamefont {Ralph},\ and\ \citenamefont {White}}]{Branczyk2011}%
  \BibitemOpen
  \bibfield  {author} {\bibinfo {author} {\bibfnamefont {A.~M.}\ \bibnamefont
  {Bra\'{n}czyk}}, \bibinfo {author} {\bibfnamefont {A.}~\bibnamefont
  {Fedrizzi}}, \bibinfo {author} {\bibfnamefont {T.~M.}\ \bibnamefont {Stace}},
  \bibinfo {author} {\bibfnamefont {T.~C.}\ \bibnamefont {Ralph}}, \ and\
  \bibinfo {author} {\bibfnamefont {A.~G.}\ \bibnamefont {White}},\ }\bibfield
  {title} {\enquote {\bibinfo {title} {Engineered optical nonlinearity for
  quantum light sources},}\ }\href {\doibase 10.1364/OE.19.000055} {\bibfield
  {journal} {\bibinfo  {journal} {Opt. Express}\ }\textbf {\bibinfo {volume}
  {19}},\ \bibinfo {pages} {55--65} (\bibinfo {year} {2011})}\BibitemShut
  {NoStop}%
\bibitem [{\citenamefont {Reichert}, \citenamefont {Defienne},\ and\
  \citenamefont {Fleischer}(2018)}]{Reichert2018}%
  \BibitemOpen
  \bibfield  {author} {\bibinfo {author} {\bibfnamefont {M.}~\bibnamefont
  {Reichert}}, \bibinfo {author} {\bibfnamefont {H.}~\bibnamefont {Defienne}},
  \ and\ \bibinfo {author} {\bibfnamefont {J.~W.}\ \bibnamefont {Fleischer}},\
  }\bibfield  {title} {\enquote {\bibinfo {title} {Massively parallel
  coincidence counting of high-dimensional entangled states},}\ }\href
  {\doibase 10.1038/s41598-018-26144-7} {\bibfield  {journal} {\bibinfo
  {journal} {Scientific Reports}\ }\textbf {\bibinfo {volume} {8}},\ \bibinfo
  {pages} {2045--2322} (\bibinfo {year} {2018})}\BibitemShut {NoStop}%
\bibitem [{\citenamefont {Zhang}\ \emph {et~al.}(2020)\citenamefont {Zhang},
  \citenamefont {England}, \citenamefont {Nomerotski}, \citenamefont {Svihra},
  \citenamefont {Ferrante}, \citenamefont {Hockett},\ and\ \citenamefont
  {Sussman}}]{Zhang2020}%
  \BibitemOpen
  \bibfield  {author} {\bibinfo {author} {\bibfnamefont {Y.}~\bibnamefont
  {Zhang}}, \bibinfo {author} {\bibfnamefont {D.}~\bibnamefont {England}},
  \bibinfo {author} {\bibfnamefont {A.}~\bibnamefont {Nomerotski}}, \bibinfo
  {author} {\bibfnamefont {P.}~\bibnamefont {Svihra}}, \bibinfo {author}
  {\bibfnamefont {S.}~\bibnamefont {Ferrante}}, \bibinfo {author}
  {\bibfnamefont {P.}~\bibnamefont {Hockett}}, \ and\ \bibinfo {author}
  {\bibfnamefont {B.}~\bibnamefont {Sussman}},\ }\bibfield  {title} {\enquote
  {\bibinfo {title} {Multidimensional quantum-enhanced target detection via
  spectrotemporal-correlation measurements},}\ }\href {\doibase
  10.1103/PhysRevA.101.053808} {\bibfield  {journal} {\bibinfo  {journal}
  {Phys. Rev. A}\ }\textbf {\bibinfo {volume} {101}},\ \bibinfo {pages}
  {053808} (\bibinfo {year} {2020})}\BibitemShut {NoStop}%
\end{thebibliography}%

\end{document}


\preprint{}

\title[Supplementary Information]{Designing High-Power, Octave Spanning Entangled Photon Sources for Quantum Spectroscopy - Supplementary Information}

\author{S. Szoke}
\affiliation{ 
Division of Engineering and Applied Sciences, California Institute of Technology, Pasadena, CA 91125, USA
}%

\author{M. He}
\author{B.P. Hickam}
\author{S.K. Cushing}
 \email[]{scushing@caltech.edu}
\affiliation{%
Division of Chemistry and Chemical Engineering, California Institute of Technology, Pasadena, CA 91125, USA
}%

\date{\today}

\maketitle

\section{\label{Theory}Theory}

\subsection{\label{refractive-index}CLT Refractive Indices}

The Sellmeier equations \cite{Moutzouris2011} describing the frequency and temperature dependent refractive indices of the material are plotted in Fig.\ref{refractive-index}.

\begin{figure}[ht]
\includegraphics[scale=0.5]{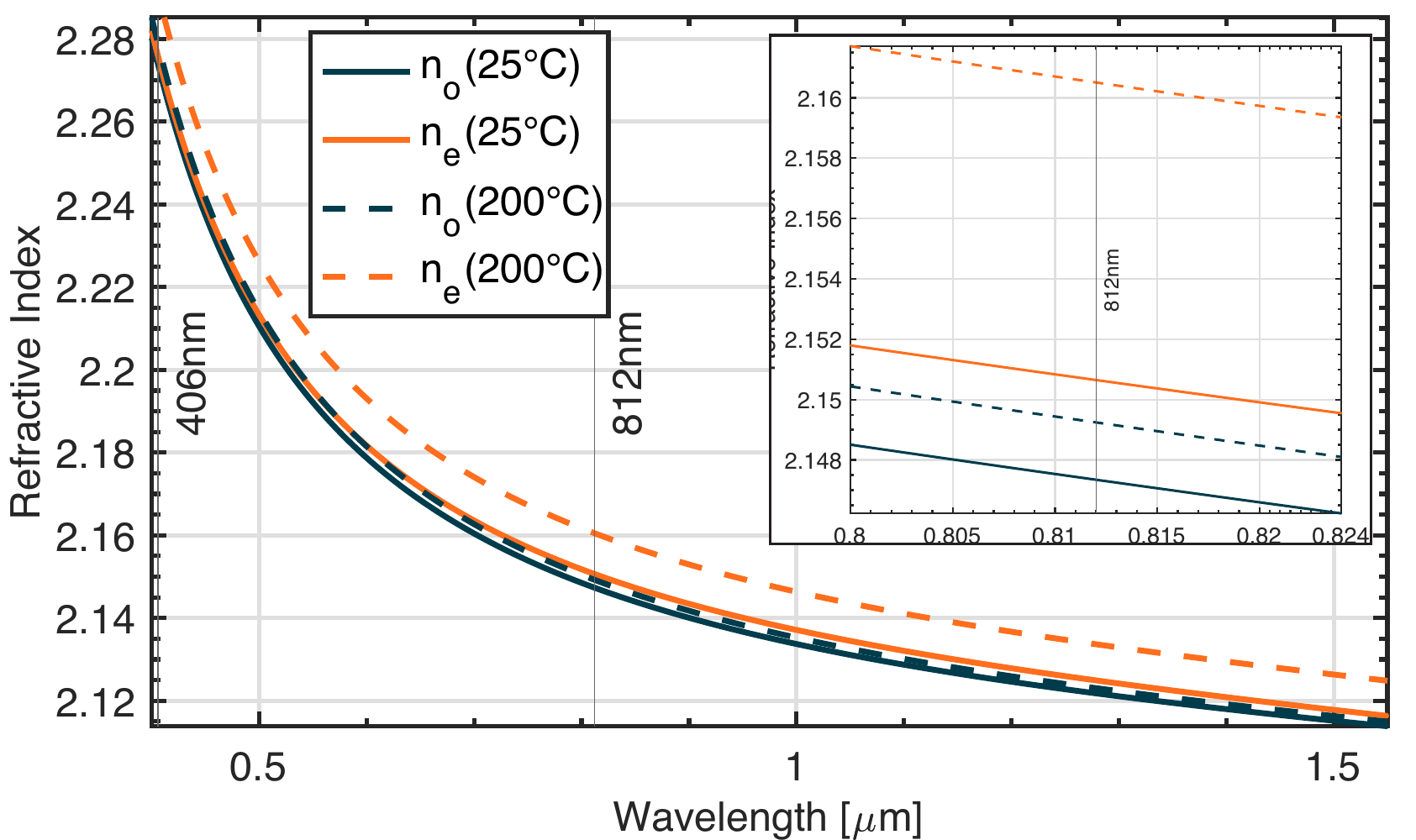}%
\caption{Congruent lithium tantalate refractive indices for ordinary and extraordinary polarizations at two different temperatures.\label{refractive-index}}%
\end{figure}

\subsection{\label{phase-matching}Quasi-Phase-Matching}

Based upon these values, the poling periodicity was calculated for a collinear $3^{rd}$-order Type-0 QPM configuration with a degenerate wavelength of 812nm.

\begin{figure}[ht]
\includegraphics[scale=0.5]{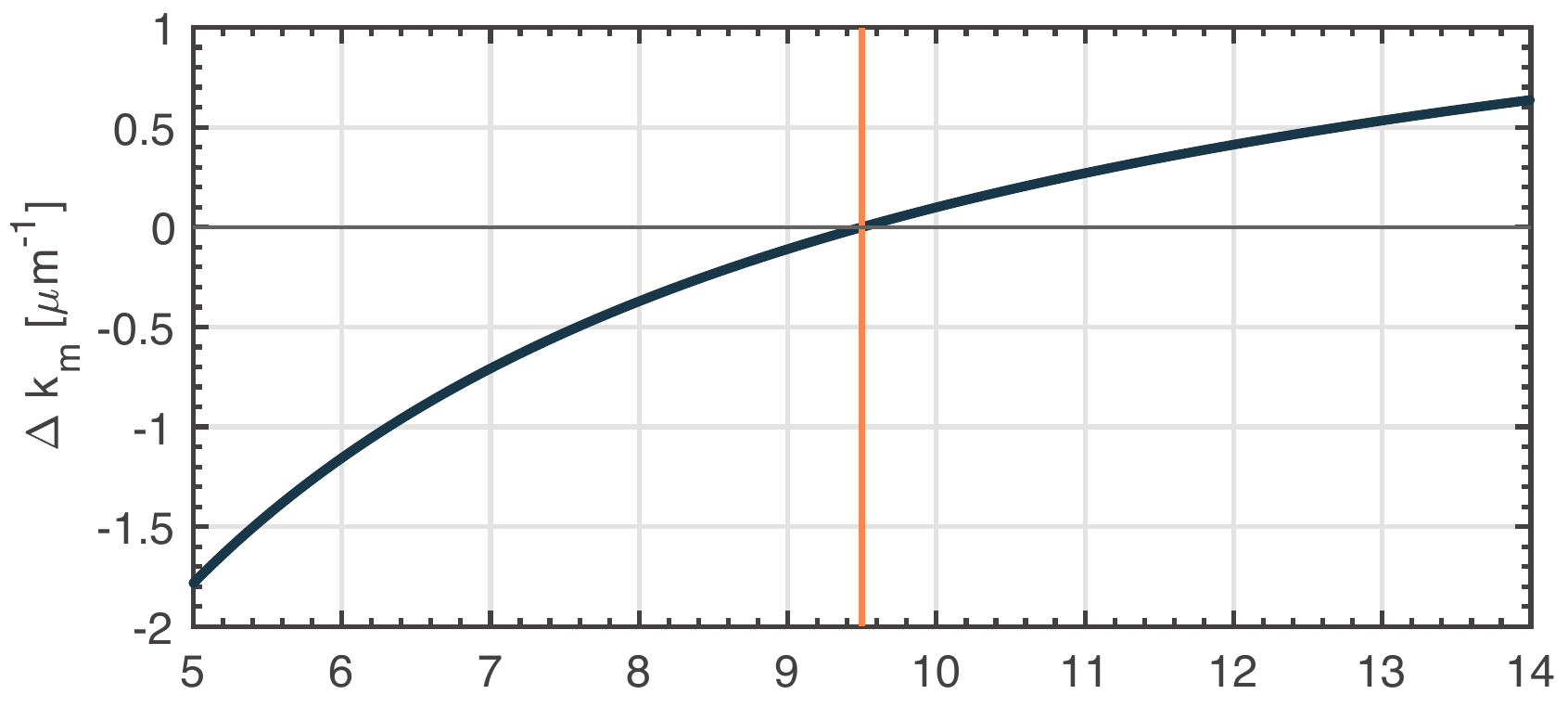}%
\\
\includegraphics[scale=0.5]{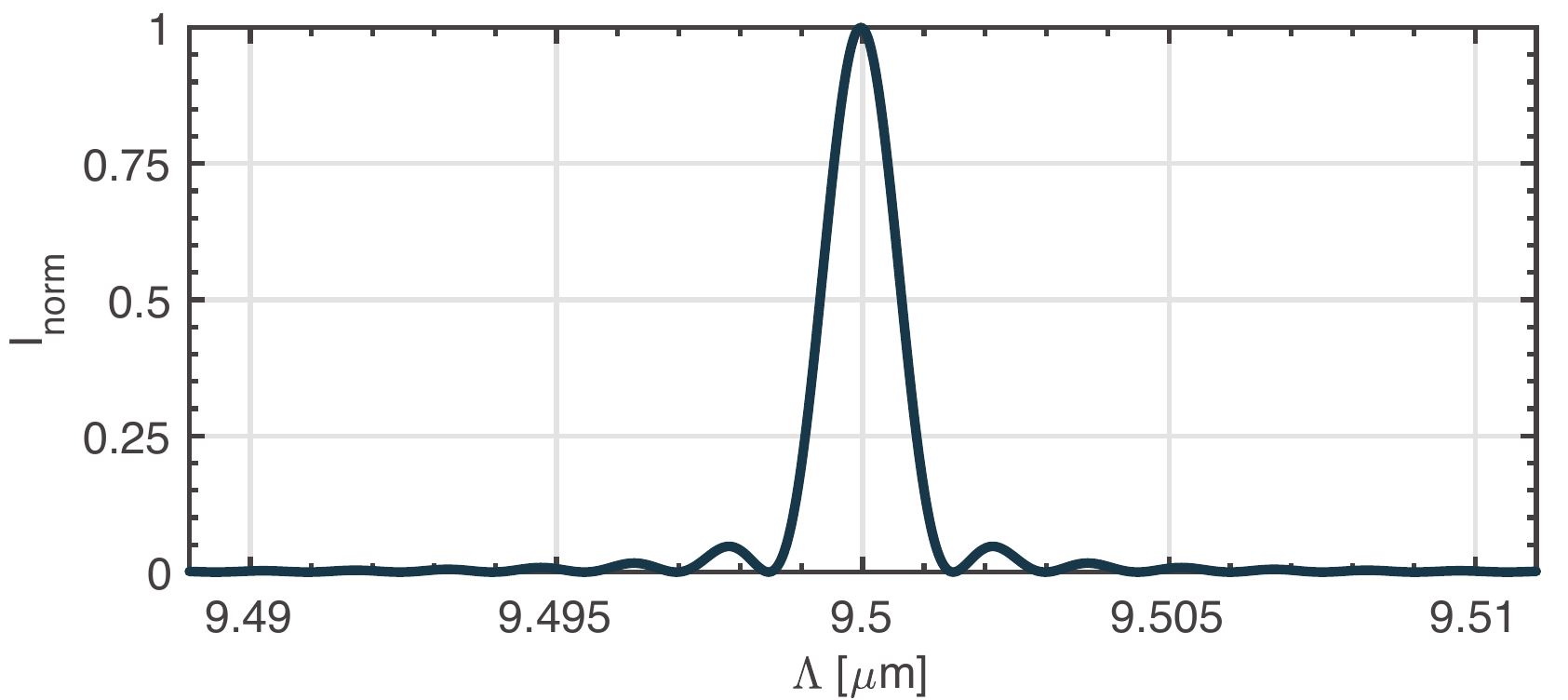}%
\caption{calculated poling period for 406nm-812nm at 133\degree C - CLT\label{CLT-poling-period}}%
\end{figure}

\subsection{\label{phase-matching}Phase-matching Amplitude}

\begin{figure}[h!]
\includegraphics[scale=0.6]{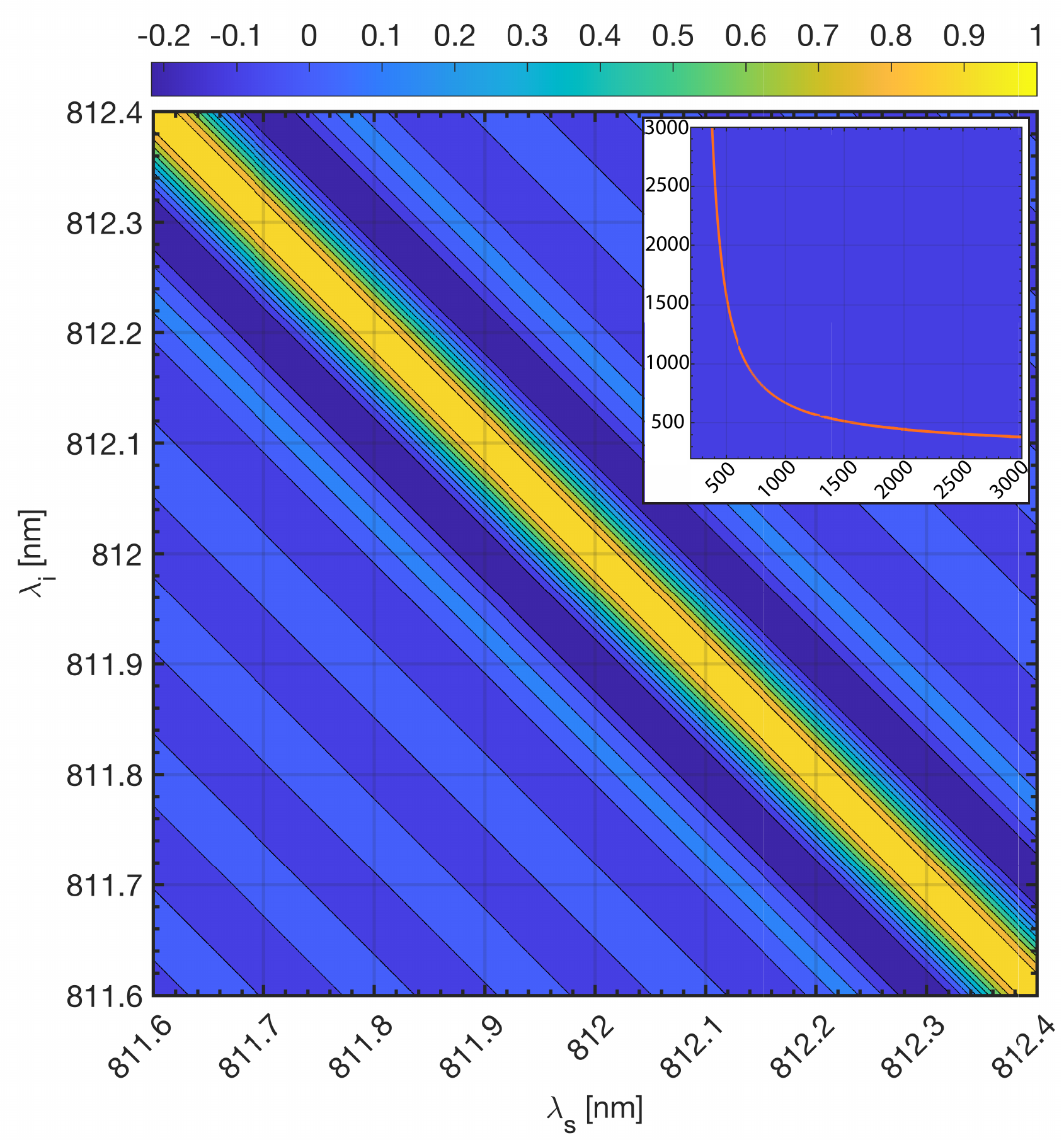}%
\caption{Phase matching amplitude contour plot for a 20 mm long ppCLT grating with a 9.5 $\mu$m poling period at 133C. Inset showing wide range phase matching curve across a larger wavelength range for the Type-0 phase matching configuration.\label{phasematching-contour}}%
\end{figure}

Once the required poling period for our specific crystal and configuration has been determined, the two-dimensional phase-matching amplitude can be calculated by again evaluating Eq.5 and Eq.7 in the main text for a wide range of wavelength combinations. The result is the contour plot in Fig.\ref{phasematching-contour}, which shows the downconversion intensity for photon pairs near the 812 nm degeneracy point. The inset of Fig.\ref{phasematching-contour} is a wide-range plot displaying the characteristic shape of the phase-matching curve for a Type-0 process. The contour provides some important information regarding the resultant output of the SPDC process. Mathematically, the output state of the downconversion process can be described as:
\begin{equation}
    \ket{\Psi}=\int\int\gamma \left ( \omega_s,\omega_i \right )d\omega_s d\omega_i a^\dagger_s a^\dagger_i\ket{0}
    \label{SPDC-state}
\end{equation}
whereby the \textit{joint spectral amplitude} (JSA), $\gamma \left ( \omega_s,\omega_i \right )$ is the product of two terms, the \textit{pump-envelope function} $\mu\left(\omega_s,\omega_i\right)$ and the \textit{phase-matching amplitude} $\phi\left(\omega_s,\omega_i\right)$ \cite{Zielnicki2018}. Here the phase-matching amplitude function is solely determined by the crystal and the desired downconversion configuration. The pump-envelope function on the other hand is dictated by, as the name implies, the spectral characteristic of the pump beam. Obviously due to conservation of energy, $\mu$ will be diagonal in its orientation if plotted as a function of the signal and idler wavelengths.

The strongly diagonal direction of $\phi$ is advantageous as its overlap with the pump-envelope along the diagonal allows for a wide range of anti-correlated signal and idler photon pairs to be phase-matched. This is of particular importance in experiments where this anti-correlation is necessary for achieving two-photon absorption processes which occur via an intermediate energy level transition \cite{Saleh1998}. This can be viewed in contrast to applications where highly spectrally pure photon states are desired, such as in single photon sources. Here, obtaining a lack of frequency correlations by designing an anti-diagonal direction of $\phi$ is of primary concern.

\subsection{\label{entanglement-purity}Entanglement and Purity}

\begin{figure}[ht]
\includegraphics[scale=0.6]{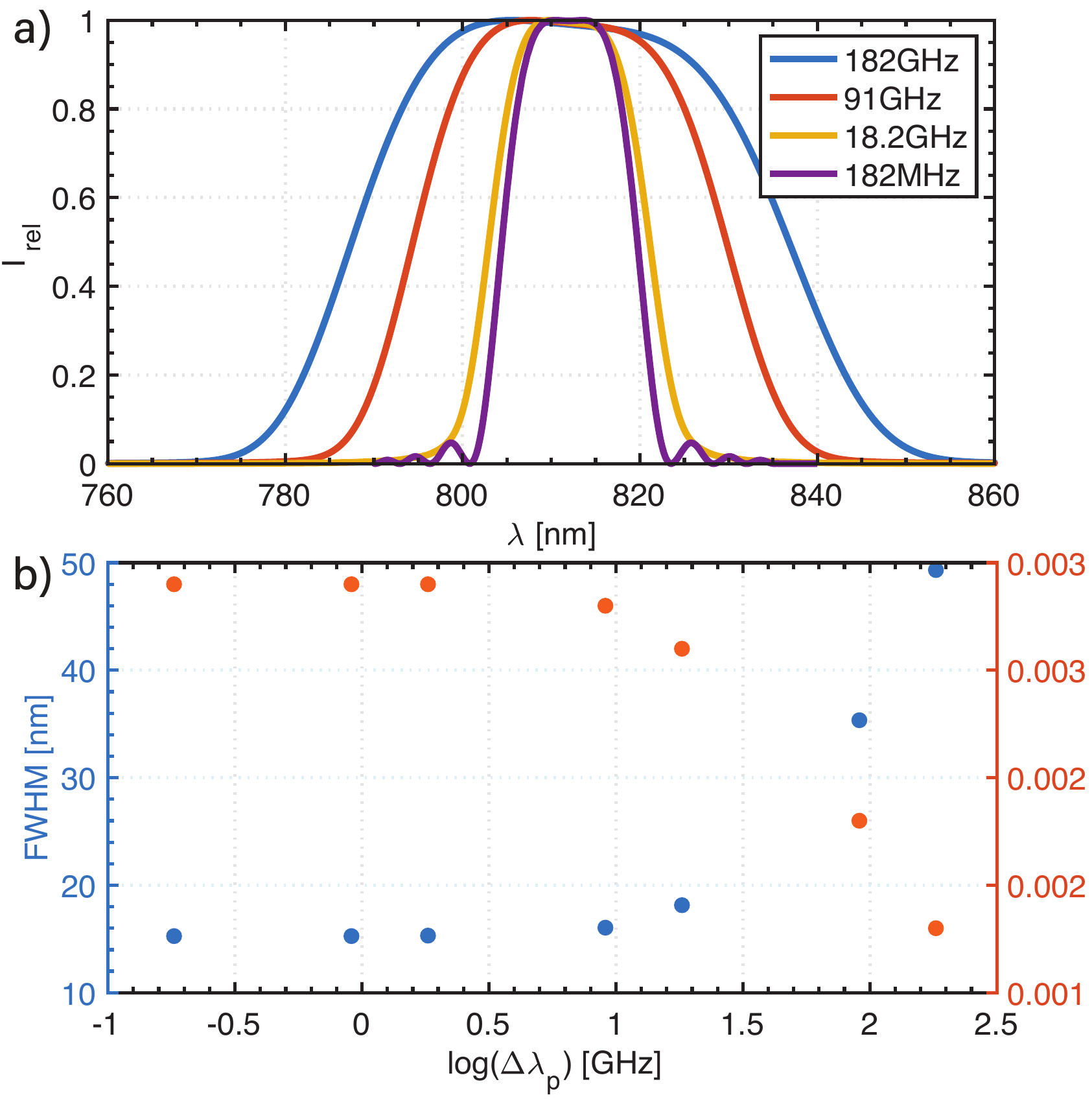}%
\caption{SPDC emission spectrum as a function of the pump bandwidth. The corresponding broadening of the spectrum at larger bandwidths carries with it a degradation of the photon state purity which is of advantage where strong (anti-)correlation between photon pairs is desired\label{ppCLT-spectrum_pumpBW}}%
\end{figure}

Given the output state as defined in Eq.\ref{SPDC-state}, the joint spectral amplitude allows for the prediction of the SPDC spectrum obtained for a given configuration of $\mu$ and $\phi$ by carrying out an integral across all signal/idler wavelengths with respect to a corresponding idler/signal wavelength held constant. For our specific 20 mm long ppCLT grating with a 9.5 $\mu$m poling period, the pump bandwidth has a distinct effect on not only the resulting emission spectrum's bandwidth, but also more profoundly on the entanglement of the photonic state. This is characterized by the purity,
\begin{equation}
    P=\frac{1}{K}
\end{equation}
with
\begin{equation}
    K=\frac{1}{\Sigma_i\lambda_i^2}
\end{equation}
being the Schmidt number equal to the rank of the reduced single particle density matrix. This value can be interpreted as a weighted measure (weights being the eigenvalues $\lambda_i$) of the number of modes required to represent the mixed state, and therefore how entangled it is \cite{Knight1995}. The eigenvalues themselves are obtained via a singular value decomposition of the joint spectral amplitude matrix. As shown in Fig.\ref{ppCLT-spectrum_pumpBW} (top), the SPDC spectrum trivially narrows as the pump bandwidth is decreased up to a limiting point. Below this value, the intersection of the pump envelope $\mu$ with the phase matching amplitude $\phi$ no longer causes a decrease in the spectral width which is being carved out by $\mu$. Fig.\ref{ppCLT-spectrum_pumpBW} (bottom) shows the calculated FWHM of the SPDC spectrum as a function of the pump bandwidth, as well as the corresponding purity of the output state. As is evident, the purity of the state decreases with increasing pump bandwidths, and therefore conversely the entanglement of the overall state increases. While this is advantageous from the perspective of producing a state which displays more/stronger anti-correlations between photon pairs, the significant drawback is that the spectral resolution with which optical atomic \& molecular transitions are able to be addressed is directly determined by the bandwidth of the pump laser driving the SPDC process. Thus, for highly spectrally resolved measurements, the figure of merit will be dictated by the pump laser.

\subsection{\label{HOM}Broadband HOM Interference}

The temporal properties and the degree of entanglement of the SPDC flux can be assessed by measuring the fourth-order interference in a two-photon Michelson (or Mach-Zehnder) interferometer. In such an interferometer, six beamsplitter ports are involved. See diagram below (Fig.\ref{michelson-diagram}).

\begin{figure}[h]
    \includegraphics[scale=0.5]{Images/Michelson MZ diagram.png}
    \centering
\caption{Schematics of a two-photon Michelson or Mach-Zehnder interferometer. \label{michelson-diagram}}%
\end{figure} 

The interferometer input wavefunction for entangled pairs entering BS1 via port \textit{a} is
\begin{equation}
    \ket{\Psi} \propto \iint d\omega_s d\omega_i \psi(\omega_s,\omega_i)\hat{a}^{\dagger}_s(\omega_s)\hat{a}^{\dagger}_i(\omega_i)\ket{0,0}
\end{equation}
where \(\psi(\omega_s,\omega_i)\) is the joint spectral amplitude of the entangled photons as defined in previous sections.

After the first pass through BS1,
\begin{equation}
\begin{aligned}
    \hat{a}^{\dagger}_s(\omega_s)\hat{a}^{\dagger}_i(\omega_i) \: = \: \frac{1}{2}[\hat{c}^{\dagger}_s(\omega_s)\hat{c}^{\dagger}_i(\omega_i) 
    - \hat{d}^{\dagger}_s(\omega_s)\hat{d}^{\dagger}_i(\omega_i)
    + i\hat{c}^{\dagger}_s(\omega_s)\hat{d}^{\dagger}_i(\omega_i)
    + i\hat{d}^{\dagger}_s(\omega_s)\hat{c}^{\dagger}_i(\omega_i)].
    \end{aligned}
\end{equation}

After the second pass
\begin{equation}
\begin{aligned}
    \hat{a}^{\dagger}_s(\omega_s)\hat{a}^{\dagger}_i(\omega_i) \: = \: \frac{1}{4} \:[1 + e^{i\omega_p\tau} - e^{i\omega_i\tau} - e^{i\omega_s\tau}] \: \hat{e}^{\dagger}_s(\omega_s)\hat{e}^{\dagger}_i(\omega_i) \\
    + \: \frac{i}{4} \: [1 - e^{i\omega_p\tau} + e^{i\omega_i\tau} - e^{i\omega_s\tau}] \: \hat{e}^{\dagger}_s(\omega_s)\hat{f}^{\dagger}_i(\omega_i) \\
    + \: \frac{i}{4} \: [1 - e^{i\omega_p\tau} - e^{i\omega_i\tau} + e^{i\omega_s\tau}] \: \hat{f}^{\dagger}_s(\omega_s)\hat{e}^{\dagger}_i(\omega_i) \\
    - \: \frac{1}{4} \: [1 + e^{i\omega_p\tau} + e^{i\omega_i\tau} + e^{i\omega_s\tau}] \: \hat{f}^{\dagger}_s(\omega_s)\hat{f}^{\dagger}_i(\omega_i).
\end{aligned}
\end{equation}
Since the coincidences are detected after port \textit{f}, the last term of the above expression is relevant. Using the fact that \(\omega_i = \omega_p - \omega_s\), the \(ff\) component of the entangled two photon wavefunction becomes
\begin{equation}
\begin{aligned}
    \ket{\Psi}_{ff} \propto - \frac{1}{4} \iint d\omega_s d\omega_i \psi(\omega_s, \omega_i)[1 + e^{i\omega_p\tau}(1 + e^{-i\omega_s\tau}) + e^{i\omega_s\tau}] \:\hat{f}^{\dagger}_s(\omega_s)\hat{f}^{\dagger}_i(\omega_i)\ket{0,0}\\
    = -\sqrt{2} \int d\omega_s \psi(\omega_s, \omega_p - \omega_s) [1 + cos((\omega_p-\omega_s)\tau)] [1 + cos(\omega_s\tau)] \: \ket{0,2}
\end{aligned}
\end{equation}
Note that here it was assumed that \(\omega_p\) is constant and the pump laser has an infinitely narrow line width. Therefore, the probability of detecting coincidences after port \textit{f} is
\begin{equation}
\begin{aligned}
    P \propto 2 \int d\omega_s \lvert \psi(\omega_s, \omega_p - \omega_s) \rvert ^2 [1 + cos((\omega_p-\omega_s)\tau)]^2 [1 + cos(\omega_s\tau)]^2
\end{aligned}
\end{equation}
where \(\lvert \psi(\omega_s, \omega_p - \omega_s) \rvert ^2\) is the joint spectral density of the entangled photons.
Thus the interference pattern at port \textit{f} can be plotted out (Fig. \ref{michelson-sim-812-10} and \ref{michelson-sim-812-100}). The interference pattern is the result of contributions from both classical and quantum (HOM) interferences. The effect of HOM is manifested by the main peak at zero time delay. As the bandwidth of the entangled photons change, either by varying phase matching conditions or simply placing a bandpass filter before the photon counting detectors, the width of the interference peak, proportional to the coherence length of the flux, changes accordingly.
\begin{figure}[h]
\includegraphics[scale=0.6]{Images/812-10 michelson pub.pdf}%
\caption{Simulated normalized narrowband Michelson interference corresponding to the Gaussian SPDC spectrum filtered by 10 nm bandpass filter centered at 812 nm. Sampling step 0.32 fs. \label{michelson-sim-812-10}}
\end{figure}
\begin{figure}[h]
\includegraphics[scale=0.6]{Images/812-125 michelson pub.pdf}
\caption{Simulated normalized broadband Michelson interference corresponding to the Gaussian SPDC spectrum filtered by 125 nm bandpass filter centered at 812 nm. Sampling step 0.32 fs. \label{michelson-sim-812-100}}%
\end{figure}

\newpage
\section{\label{Data}Additional Experimental Data}

Figs.6-9 provide additional characterization on the relationship between interference measurements and the pump.

\begin{figure}[h]
\includegraphics[width=\columnwidth]{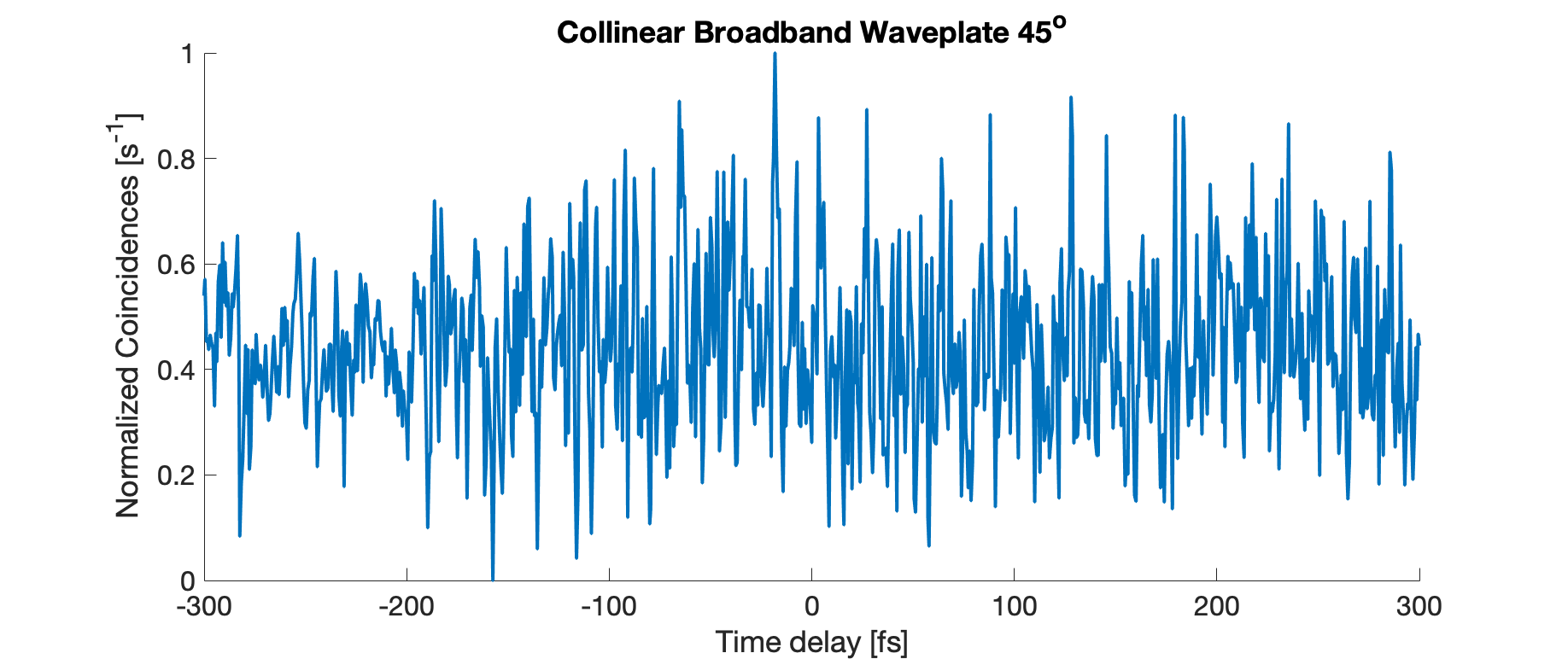}%
\caption{Measured collinear broadband two-photon interference when the arms of the Michelson interferometer are perpendicularly polarized. No distinguishable interference pattern is observed, and the coincidences follow pump and noise fluctuations.  \label{45degreebroadband}}%
\end{figure}

\begin{figure}[h]
\includegraphics[width=\columnwidth]{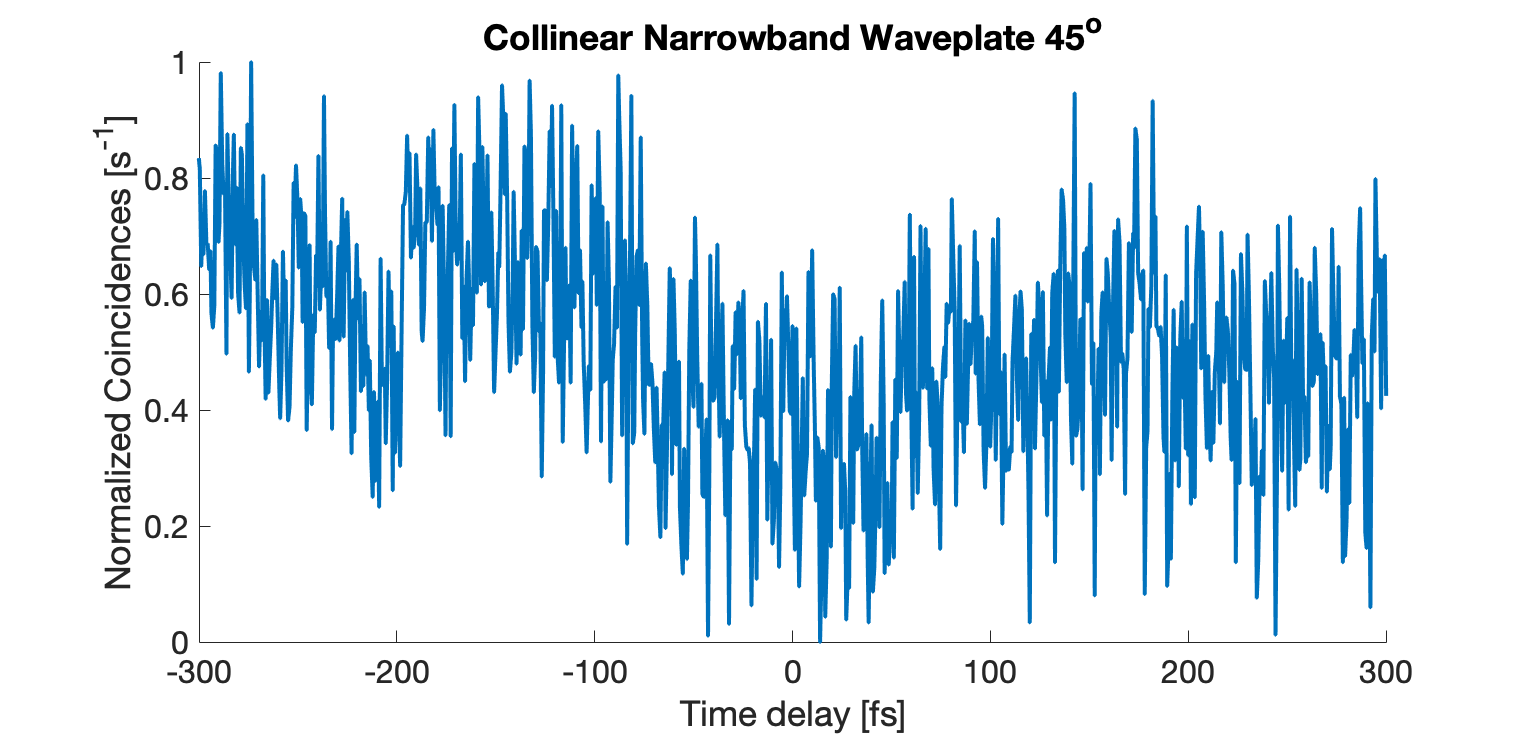}%
\caption{Measured collinear narrowband two-photon interference when the arms of the Michelson interferometer are perpendicularly polarized. No distinguishable interference pattern is observed, and the coincidences follow pump and noise fluctuations. \label{45degreenarrowband}}%
\end{figure}

\begin{figure}[h]
\includegraphics[width=\columnwidth]{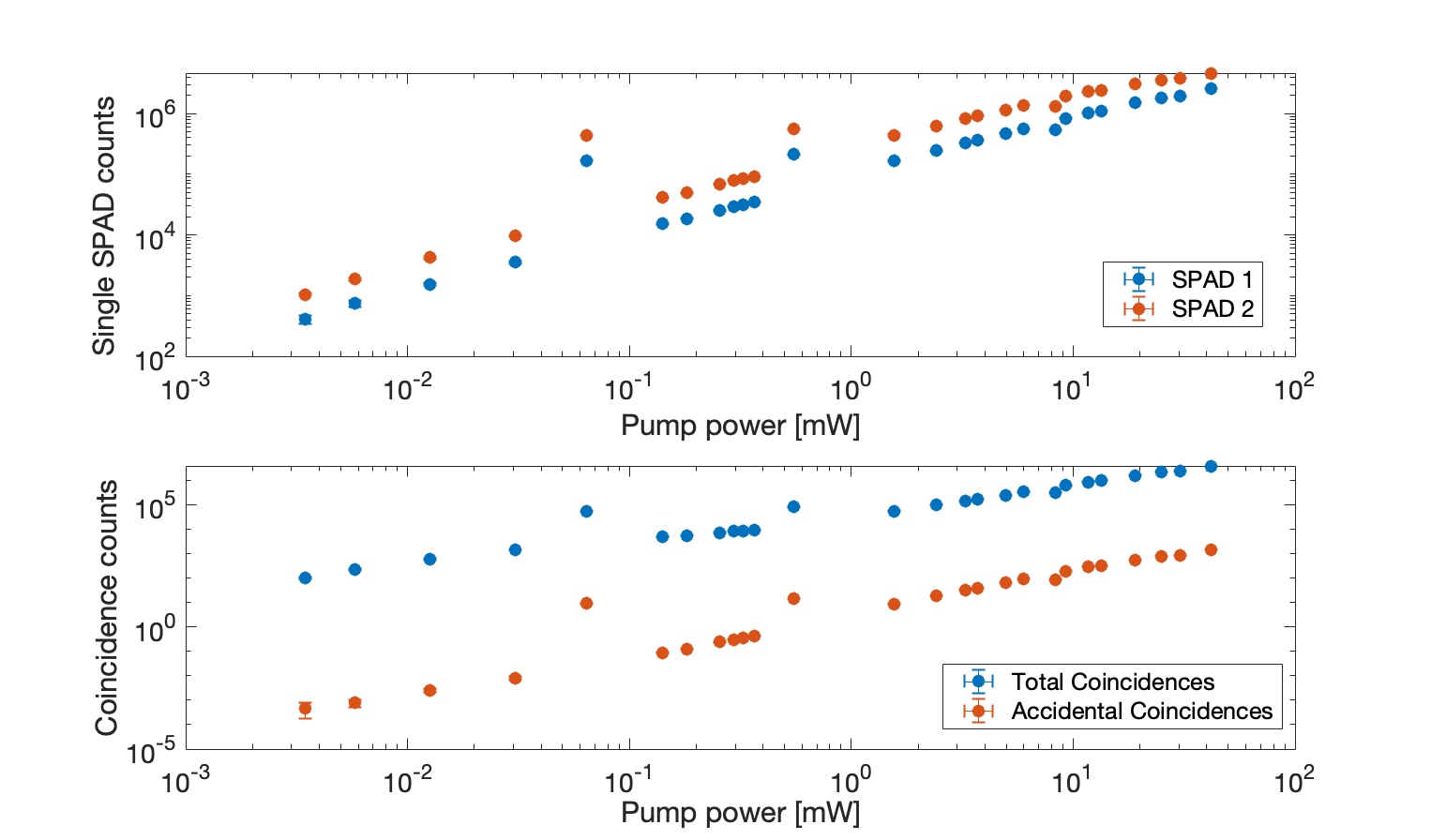}%
\caption{Measured SPAD counts and coincidences vs. pump power in the collinear broadband case. Top: single SPAD channel counts in 15s. Bottom: Total and accidental coincidences in 15s. \label{broadbandcoincidencesvspump}}%
\end{figure}
\begin{figure}[h]
\includegraphics[width=\columnwidth]{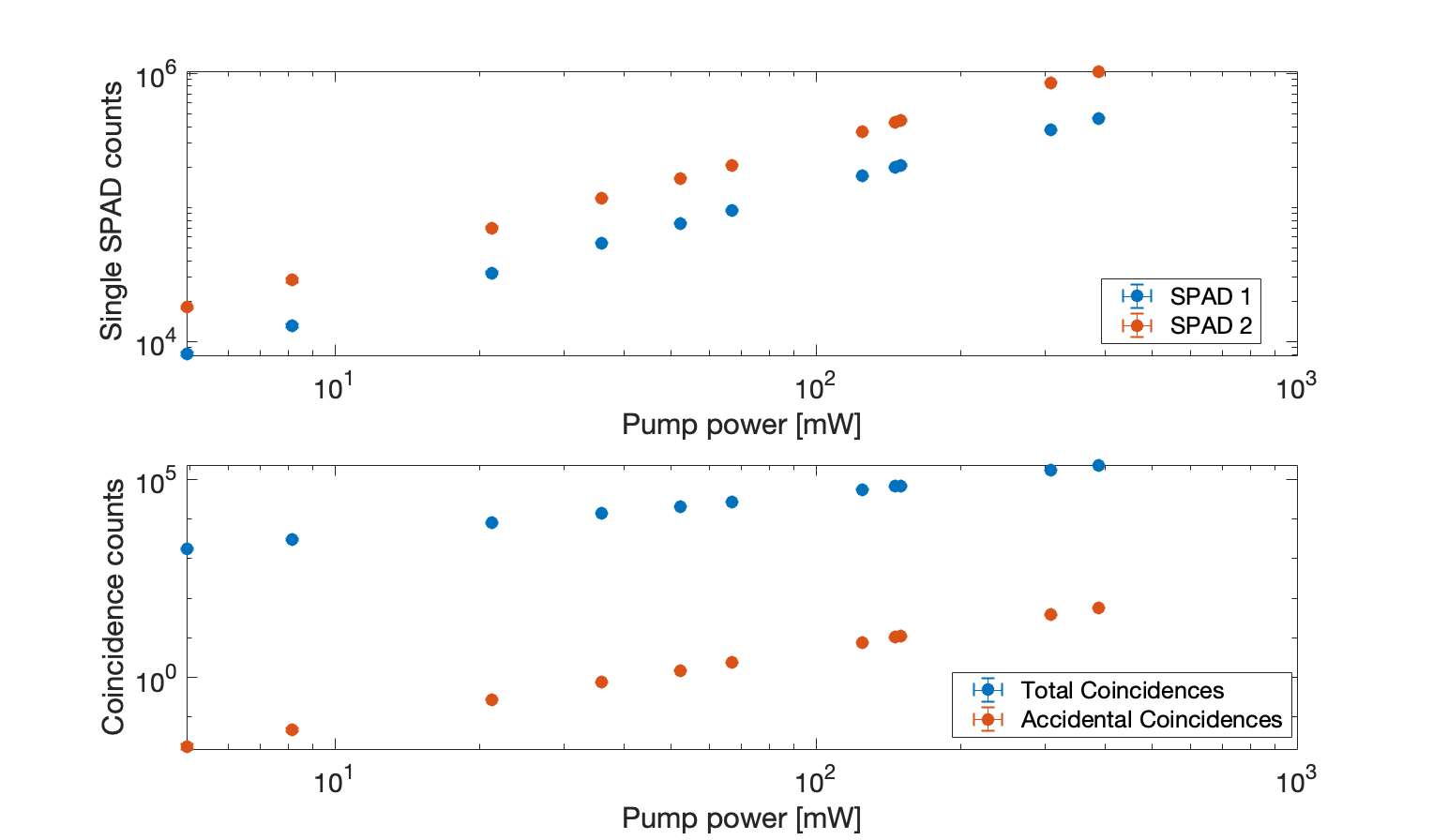}%
\caption{Measured SPAD counts and coincidences vs. pump power in the collinear narrowband case. Top: single SPAD channel counts in 15s. Bottom: Total and accidental coincidences in 15s. \label{narrowbandcoincidencesvspump}}%
\end{figure}

\clearpage
\bibliography{main.bib}